# Dialogue Enhancement in Object-based Audio

Evaluating the Benefit on People above 65

**Davide Sönke Straninger**

A thesis presented for the degree of
Bachelor of Arts (B.A.)

| | |
|---|---|
| **Degree programme:** | Multimedia and Communication |
| **University:** | Ansbach University of Applied Sciences |
| **Submitted by:** | Davide Straninger, <davide.straninger(at)gmx.de> |
| **Date:** | 13.03.2020 |
| **1. Examiner:** | Prof. Dr. Cornelius Pöpel |
| | Ansbach University of Applied Sciences, |
| | Ansbach, Germany |
| **2. Examiner:** | Dipl. Tonmeister Christian Simon |
| | Fraunhofer Institute for Integrated Circuits IIS, |
| | Erlangen, Germany |

# 0. Abstract

Due to age-related hearing loss, elderly people often struggle with following the language on TV. Because they form an increasing part of the audience, this problem will become even more important in the future and needs to be addressed by research and development. Object-based audio is a promising approach to solve this issue as it offers the possibility of customizable dialogue enhancement (DE). For this thesis an Adjustment / Satisfaction Test (A/ST) was conducted to evaluate the preferred loudness difference (LD) between speech and background in people above 65. Two different types of DE were tested: DE with separately available audio components (speech and background) and DE with components created by blind source separation (BSS). The preferred LDs compared to the original, differences of the preferred LDs between the two DE methods and the listener satisfaction were tested. It was observed that the preferred LDs were larger than the original LDs, that customizable DE increases listener satisfaction and that the two DE methods performed comparably well in terms of preferred LD and listener satisfaction. Based on the results, it can be assumed that elderly viewers above 65 will benefit equally from user-adjustable DE by available components and by dialogue separation.

# Contents

# List of Figures

# List of Tables

# List of abbreviations

| | |
|---|---|
| **A/ST** | Adjustment / Satisfaction Test |
| **AR** | Artificially created |
| **BASS** | Blind Audio Source Separation |
| **BBC** | British Broadcasting Corporation |
| **BG** | Background |
| **BSS** | Blind Source Separation |
| **CBA** | Channel-based Audio |
| **DE** | Dialogue Enhancement |
| **DRC** | Dynamic Range Control |
| **DS** | Dialogue Separation |
| **EBU** | European Broadcasting Union |
| **FG** | Foreground |
| **GUI** | Graphical User Interface |
| **IQR** | Interquartile Range |
| **ITC** | Independent Television Commission |
| **ITU** | International Telecommunication Union |
| **LD** | Loudness Difference |
| **LU** | Loudness Units |
| **MPEG-H** | MPEG-H 3D Audio |
| **MUSHRA** | Multi-Stimulus Test with Hidden Reference and Anchor |
| **NGA** | Next Generation Audio |
| **OBA** | Object-based Audio |
| **OO** | Original Objects |
| **P-ID** | Participant-Identifier |
| **SNR** | Signal-to-Noise Ratio |
| **WDR** | Westdeutscher Rundfunk |
| **WHO** | World Health Organization |

# Acknowledgement

I would like to thank Prof. Dr. Cornelius Pöpel, who equipped me with the knowledge to explore the field of audio research and who made this work possible in the first place. Furthermore I would like to thank Christian Simon, who supported and motivated me for the topic of accessibility in broadcasting. Special thanks also go to Matteo Torcoli, for his support with the statistical analysis of the results, for always having an open ear and for his great patience. Thanks also go to Dr. Jouni Paulus for the support with his vast knowledge and to the Fraunhofer IIS listening test team for their help and experience in dealing with test subjects. Finally, I would also like to thank my family and friends who understood my lack of time.

# 1. Introdcution

"It is not the voice that commands the story: it is the ear."
(Italo Calvino, Invisible Cities)

## 1.1. Problem description

Television is a cultural asset and an important information medium. In order to participate in societal developments, it is therefore important to make the language on television understandable to as many people in society as possible. Since demographics in almost all modern societies are changing towards a higher average age, and increasing age is often accompanied by hearing loss, there is a need to improve speech intelligibility in television.
Modern audio systems are capable of improving the needs of a more and more diverse and ageing audience. Especially Dialogue Enhancement technologies in combination with object-based audio seem to have high potential of providing solutions for various challenges.

## 1.2. Motivation for the topic

The field of audio science is vast and offers plenty of room for new ideas, findings and inventions. But it is usually the basic needs that are easily overlooked and left behind. This also applies to the much criticised topic of speech intelligibility. But with the latest technical innovations and developments there are now unprecedented possibilities to meet the individual needs of people and to address existing problems. With this work, I wish to contribute to societal improvements, to enable people with age-related hearing impairment to maintain their feeling of independence, while also facilitating their participation in art and culture.

## 1.3. Research question and scope of the work

The purpose of this work is to view, discuss, collate and relate the theoretical basics of age-related hearing loss and dialogue enhancement in relation to object-based audio. A listening test will be conducted. It is firstly intended to evaluate if the possibility of having a personifiable dialogue enhancement through an adjustable loudness difference between speech and background (audio components) with dialogue separation or available audio components causes an improvement on people above 65, regarding the overall listener satisfaction. Secondly it shall evaluate if there are differences between dialogue-separation-based and stem-based dialogue enhancement noticeable for elderly viewers and what their preferred loudness differences are.

This work makes no claim to completion and can also not take into account the full range of age-related sensory, physical and mental impairments due to the limited extent of a Bachelor thesis.[1]

---

[1] Due to the anonymization of the test participants and for the concurrent simplification of the text flow, the generic masculine is always used in this work. Whenever a gender-specific term is used, it should be understood as referring to all genders, unless explicitly stated.

# 2. Theoretical basics

For understanding why accessibility of TV and broadcast audio is important and how the problem of poor speech intelligibility can be solved, it is necessary to first take a look at some theoretical basics concerning the topics age-related hearing loss, Dialogue Enhancement (DE) and Object-based Audio (OBA).

## 2.1. Hearing Loss

For solving a problem, the working principle of the defective system has to be understood to be able to analyse the malfunction itself. Transferring to human hearing, this chapter will take a closer look on how sound transmission through the auditory anatomy works and where changes in this transmission occur as age progresses. Lastly there will be given facts about the prevalence of age-related hearing loss and about television consume of elderly people.

### 2.1.1. Anatomy of the auditory system and sound transmission

The auditory system consists of two greater parts. The sensory organs (outer, middle and inner ears) and the sensory system (nerves and brain areas) which process the electric impulses received from the ears. This chapter will give a rough overview of the sound transmission through these two main structures.
As sound travels across the air, it reaches the outer ear which consists of a cartilaginous structure: the auricle or pinna. Its shape varies greatly between individuals. The sound waves are fractured according to its structure and travel further into the ear through the auditory canal (external auditory meatus) which incidentally serves as an amplifier for incoming acoustic signals. These before mentioned structures build the outer ear. The sound waves get manipulated while travelling through the ear canal and finally hit the tympanic membrane, causing it to vibrate. Connected to the membrane there is a structure of three small bones that are set into motion by the vibrations. They are called malleus, incus and stapes. The tympanic membrane and the bones together form the middle ear. Its main function is to convert the acoustic waves into mechanical vibrations and it also limits human hearing to a frequency range of about 20 Hz to 20 kHz and a range of 0 to about 130 phon.
The bone stapes directs the now mechanical vibrations onto the so called oval window, a membranous structure, witch it is connected to. Behind that window, there is a cavity that is filled with a fluid in which the mechanical waves are transmitted. The fluctuations in the liquid travel along an up-winding canal (scala vestibuli) to the top of the cochlear, a superordinate winding spiral structure. There is a second canal that winds up in the cochlea, the cochlear duct. It is in touch with the scala vestibuli through the Reissners membrane. The cochlear duct is also filled with a liquid and contains the tectorial membrane which closely covers the hair cells. Through the basilar membrane, the cochlear duct touches a third fluid-filled canal that is winding down the cochlear. It is called the scala timpani. So, when waves travel along the up-winding scala vestibuli, the fluid in the cochlear duct begins to vibrate and transmits these vibrations through the basilar membrane into the down-winding scala timpani. The hair cells, which sit in the basilar membrane, also begin to move up and down, causing the outer hair cells to bend against the tectorial membrane. This evokes electrical impulses inside the inner hair cells which play a major role in the conversion of mechanical waves into neural signals. These electric impulses get transmitted to the brain through the vestibular nerve. The hair cells which are closer to the entrance of the cochlear spiral respond to higher frequencies, whereas the hair cells at the top of the cochlea respond to the lower frequencies. The vibrations in the down-winding scala timpani travel all the way to a second membranous window. This is named the round window. It serves as an equalizer within the cochlea, e.g. expanding when pressure is being built up by the bone stapes, pressing on the oval window. Travelling up the primary auditory pathway, the electrical signal gets transmitted to the auditory cortex. Inside this brain region the received signals get handed over to the primary auditory cortex. In this area, there are specialized neurons that only respond to electric impulses of single acoustic frequencies. They are organized in a tonotopic order which means that their spatial position within the brain is relative to the frequency at which they become active [31, 32].

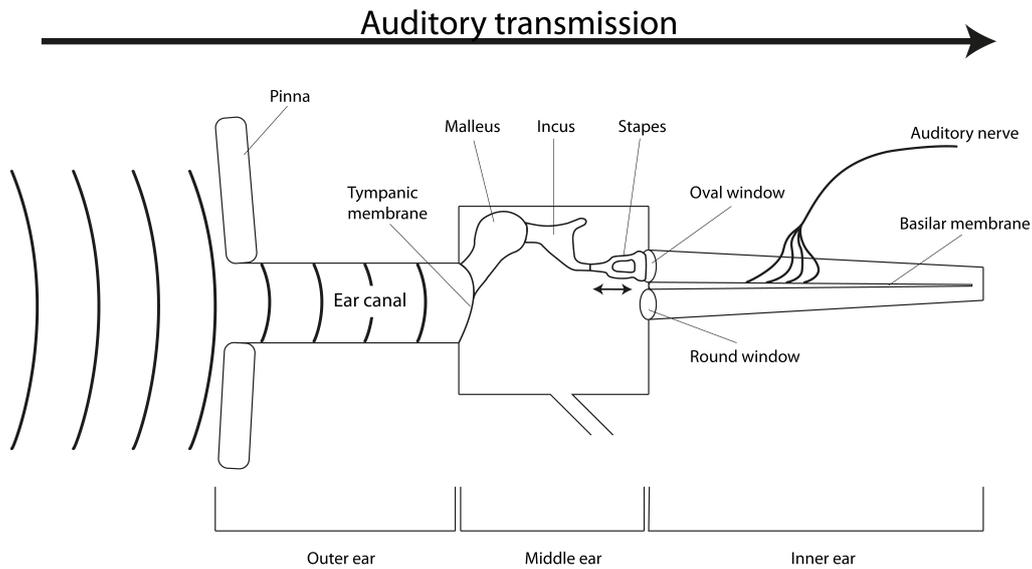

Figure 1: The schematic auditory transmission based on [25, p. 16].

### 2.1.2. Age-related changes of the auditory system and resulting hearing impairment

Hearing loss can be caused by various reasons. A congenital defect, exposing the ear to loud sound stimuli for too long, an accident, etc. This chapter focusses on hearing loss that comes with proceeding age.
Some sources claim that hearing-loss does not come "naturally" with age. Age seems to be no safe indicator for hearing impairment and there is no consistent explanation for age-related hearing loss, it just statistically happens more often in elderly people. Age and the loss of hearing ability are indirectly temporal, but not causally connected to each other. However, age-related changes of the inner-ear, the nervous structures and the areas of the brain that are in charge of processing auditory data, have been proven and are measurable. This slow decline of the auditive sensitivity of the human auditory system is also referred to as presbycusis [17].
Slowly beginning from the age of around fifty, without any apparent cause, the hearing ability gets successively worse mainly due to morphological and functional changes of the cochlea and the auditory nerve [28]. But changes can occur in all parts of the auditory transmission chain which makes age-related hearing loss multifactorial. The so-called conductive hearing loss can originate from defects in the outer and middle ear, sensory hearing loss from the inner ear, especially the organ of corti; neural hearing loss describes defects in the auditory nerve, and finally central hearing loss which relates to age-related changes in specific brain areas and nodal points that are vital to the process of hearing [55].
Conductive hearing loss can be caused by various reasons like e.g. a blocked auditory canal, a damaged tympanic membrane, fluids in the middle ear or injured ossicles, but these anomalies are not significantly measurable age-related and can also occur in younger people [17]. Sensory hearing loss is one of the most prominent causes for auditory difficulties in elderly people. The outer hair cells of the organ of corti in the basilar membrane are then damaged and can therefore not stimulate the inner hair cells properly any more. This leads to a successively more imprecise representation and resolution of frequencies on the basilar membrane. The hearing threshold of elderly people commonly shows a raising characteristic in the whole frequency spectrum and significant loss in high frequencies starting from about 4kHz. A typical observed pure-tone audiogram in presbycusis is shown in figure 2.

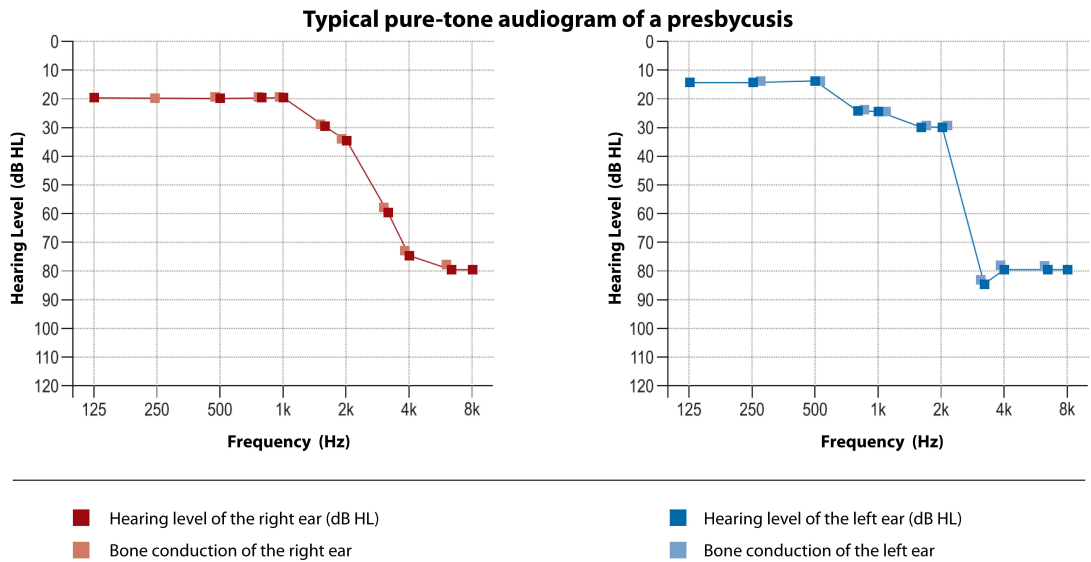

Figure 2: A typical expected pure-tone audiogram of a presbycusis for the right and left ear on air and bone conduction reproduced from [28, p. 304] with kind permission of Deutsches Ärzteblatt.

The more inner and outer hair cells are defected, the more linear the hearing gets [11, 17]. Neural hearing loss refers to problems with signal transmission in the auditory nerve and its synapses. Tumours and inflammatories around that area are among the most common reasons for this type of hearing loss [61].
Lastly there also is central hearing loss. "The higher the neural dysfunction is located, the more complex the hearing disturbance; thus, the patient may have difficulty in recognizing certain signals amid acoustic noise, in disentangling simultaneous speech signals, or in recognizing timbre." [61, p. 442]. Due to the ageing process, neurological defects occur. Neurons of e.g. the primary auditory pathway degrade and the amount of neurotransmitters as well as neuro-receptors decreases [17]. In addition to that, if the central auditory pathway does not get enough stimuli for a prolonged period of time, the ability of speech perception and processing in the brain decreases more and more [48]. A summary of the different types of hearing loss, categorized by their occurrence, can be seen in figure 3.

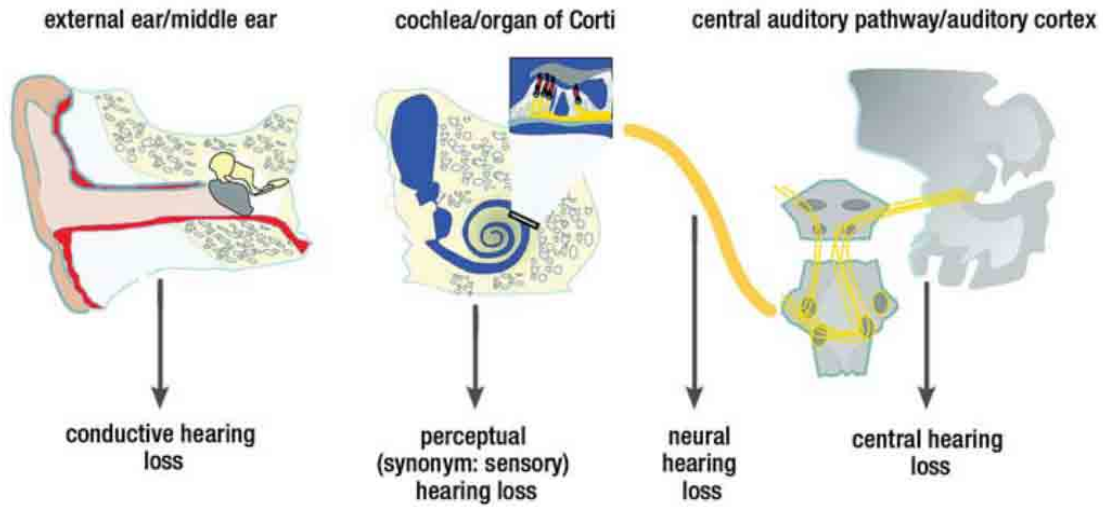

Figure 3: Topographic-functional classification of hearing impairment reproduced from [61, p. 435] by courtesy of Deutsches Ärzteblatt.

Language comprehension and understanding is not only achieved by the sensory bottom-up processes, but also the top-down mechanisms like cognition. Brain regions that are in charge of the short-time and long-time memory slowly degrade with age and the cognitive processing speed slows down [28, 55]. These factors are all important for speech perception and comprehension and therefore worsen with age. But since presbycusis is a multifactorial process it often is caused by a combination of defects like e.g. a symmetrical sensorineural loss accompanied by a loss of higher frequencies [11].
As shown above, age-related hearing loss can be categorized by the defects that cause it. Another perspective takes the problems and restrictions for affected people into account. The World Health Organization (WHO) categorized hearing loss as seen in table 1.

| Grade of hearing impairment | Audiometric ISO value (at 0.5, 1, 2, 4 kHz) | Qualitative description |
|---|---|---|
| No impairment | $\leq$ 25 dBHL (better ear) | No or very slight hearing problems. Able to hear whispers. |
| Slight/mild impairment | 26-40 dBHL (better ear) | Able to hear and repeat words spoken in normal voice at 1 meter. |
| Moderate impairment | 41-60 dBHL (better ear) | Able to hear and repeat words spoken in raised voice at 1 meter. |
| Severe impairment | 61-80 dBHL (better ear) | Able to hear some words when shouted into better ear. |
| Profound impairment or deafness | $\geq$ 81 dBHL (better ear) | Unable to hear and understand even a shouted voice. |

Table 1: WHO classification of the severity of hearing impairment based on [61, p. 434], [60].

Speech intelligibility should be a central criteria of the evaluation of hearing loss, because it is speech and communication that connects us to other people and ultimately to society. But in pure-tone-audiometry, the loss of higher frequency-perception in progressing age has only small influence on the rating. The sensorineural changes of age-related hearing loss also results in a decrease of the ability of recognizing the temporal changes in sound which is especially important for the perception and interpretation of fast changes in sound, as they appear in spoken sentences. The perception of pitch also worsens, which is a significant problem, because it contains information about contextual significance, emphasis or structure of sentences but also about age, sex and emotions of the speaker. Furthermore many studies show that spectral masking is part of sensorineural hearing loss. Elderly people with mild hearing loss normally do not have bigger problems with following what is being said, but when the background noise-level increases or the listener is located in a rather reverberant room or multiple speakers are speaking at the same time, people with hearing loss tend to have severe problems with understanding speech [11]. The difficulties of speech intelligibility and background noise in a conversation situation can easily be related to the difficulties that occur when trying to understand dialogue in TV with music, sound effects or atmospheric noise in background.
Especially nowadays, as the demands for communication increase, the disability of understanding speech correctly can lead to drastic negative consequences for elderly people. Bad speech intelligibility can finally result in a limitation of communication, social isolation, sensory deprivation and frustration up to suicidal thoughts [17, 28, 7, 48].

### 2.1.3. Prevalence of hearing impairment

It is estimated that between 2015 and 2050, the worldwide number of people aged above 60 years will double and the number of people older than 80 years will even triple. For men of the age of 60 and above, hearing loss is the number one cause, and for women in the same age span it is the second most cause of disability. So, about one third of all adults above the age of 65 are affected by hearing loss and about 226 million of these people have a disabling hearing loss. The term disabling hearing loss refers to the WHO classification of a hearing loss greater than 40 dB [60]. The number of people above 65 affected by disabling hearing loss will rise up to 585 million by 2050, as estimates show [59].

For further illustration, an average of about 18.9% of all European inhabitants are above the age of 65 and the life expectancy in the European countries will grow in the next years [45, 46]. In 2011, the proportion of the German population over 65 years of age was 21%. This is a peak value compared to all other EU countries [14]. As statistics show, in Germany the amount of TV consume per day significantly increases with age. The older people are, the more TV they watch [2]. In addition to that, many old people ignore that they have problems with hearing and understanding spoken words and often search for external causes instead of trying out hearing devices [48, 17]. This case exemplarily illustrates that the needs of an increasingly older audience will become even more relevant for TV audio in the future.

## 2.2. Dialogue Enhancement

As the previous chapter has shown, being able to follow the dialogue in broadcast audio will be of high importance in the coming years. To be able to fulfil the needs of an increasingly diverse and demanding audience it will be necessary to offer the possibility of adjusting the audio mix to the needs of the consumer. As seen in chapter 2.1.2, especially the intelligibility of speech suffers from age-related hearing loss. With the help of DE it is possible to address these demands.

### 2.2.1. What can generally be understood by Dialogue Enhancement

There is no official definition for the term of Dialogue Enhancement (DE). It generally describes all soft- and hardware techniques and methods that support and enhance speech intelligibility[2] in any audio material. A better intelligible dialogue may be given, if the Signal-to-Noise Ratio (SNR) or Loudness Difference (LD) between the Foreground (FG)[3] and Background (BG)[4] is large enough, enabling the viewer to understand what is being said. As already mentioned in chapter 2.3.2, the preference of this ratio highly depends on the type of BG (e.g. music, broadband noise or narrowband sound effects), on the listening environment, and of course on the personal hearing ability and therefore differs from one person to the other. Most institutions and manufacturers have their own names for the DE solutions they offer, such as "Hearing Boost" [39], "Dialog Clarity" [38] or "Speech Enhancement" [44]. As the previous chapters have already shown, the need for enhancing intelligibility of dialogue in broadcasting and streaming is given and will increase in the next years significantly. That some sort of DE can in general be beneficial to normal hearing and hard of hearing viewers was already shown in different studies [42, 58, 23, 8, 27]. A more in-depth look at past and present research in this field is given in chapter 3.

Enhancing the dialogue can be achieved by manual or automated processes, the latter with or without the usage of machine learning. The manual ways of improving dialogue intelligibility include for example manual ducking[5] and ducking through a side-chain-compression, or applying static compressions and filters to an audio mix. Automated processes can consist of software that automatically applies filters and compressors based on analysis of the audio mix or, for example, tools that work with databases and deep learning.

### 2.2.2. Dialogue Separation

A precondition of being able to enhance the dialogue in an already existing audio mix without the audio source material or single stems[6] being accessible is a preceding Dialogue Separation (DS). DS is referred to as the process of splitting a full audio mix into FG and BG. It is emphasized at this point that DS can not recover the original components[7] as clearly separated as they were before merging them into a full mix. Components created by a DS algorithm are only estimated FG and BG and each of the two components still contains remnants of the other. This separation can be achieved by various different automated software methods which can also include machine learning. A commonly used method for separating a signal is the so-called Blind Source Separation (BSS) or, in the context of an audio signal, also called Blind Audio Source Separation (BASS). It consists of "recovering one or several source signals from a given mixture signal" [54, p. 2] by estimating what is speech and background. A selection of well known software solutions that include DS are e.g. the RX7 by iZotope [19], the ADX SVC [6], IDC (Instant Dialogue Cleaner) and XTrax Stems 2 [5] from Audionamix or the Spleeter software [33], developed by Deezer. Most big tech companies like Samsung, Sonos or Sennheiser also offer built-in solutions for better dialogue intelligibility. This work mainly focusses on DE achieved by a DS algorithm that uses BSS in combination with machine learning.

---

[2]It is important to note that speech intelligibility is not to be equated with speech comprehension, which includes understanding what is being said and which therefore is a mainly cognitive issue. Intelligibility often refers to the proportion of words which are correctly heard and therefore is a mainly sensory issue [12, 56].

[3]This work refers to FG as speech.

[4]This work refers to BG as all ambient noise, music, sound effects and other non-speech sounds and estimations of these, done by Blind Source Separation algorithms.

[5]In this context, ducking is referred to as any time-varying background attenuation with the aim of making the foreground speech better intelligible.

[6]A stem is a pre-mix of an audio production which could e.g. consist of the dialogue, the music, the sound effects or the background ambience only.

[7]This work refers to components as FG and BG of a mix.

## 2.3. Object-based Audio

"Object-based audio is a revolutionary approach for creating and deploying interactive, personalised, scalable and immersive content, by representing it as a set of individual assets together with metadata describing their relationships and associations"[20].
OBA is described by several sources as a technology with the potential of solving current problems in broadcast and streaming audio while at the same time supporting new functions and possibilities for both broadcasters and consumers [58, 15, 10, 43]. In the following chapter, some of the basic facts about OBA and how it can improve accessibility in broadcasting will be presented.

### 2.3.1. What is Object Based Audio?

To be able to understand what OBA is, it is useful to recapitulate what Channel-based Audio (CBA) is and where the differences lie.
There is a distinction between production, transmission and reproduction methods. In conventional CBA, for each loudspeaker of the reproduction setup[8], e.g. stereo or 5.1, there has to be a dedicated channel in the audio stream of a broadcast or streaming transmission. The audio is recorded or mixed as single channels, transmitted as channels according to their dedicated reproduction format and finally presented via the reproduction system as one channel per loudspeaker. In this same way, every available language version in the programme material is also permanently mixed to these channel formats. So, e.g. if a consumer watches a documentary on a streaming platform in a certain language on stereo speakers, there will be two channels of audio transmitted with the video. If the consumer changes the language, a completely new audio stream has to be transmitted which often goes along with a new video stream, because in post-production video and audio get multiplexed. If there is the option of adding an audio description or a version with better intelligible speech, for each option a new stereo audio stream has to be transmitted.

---

[8]This work relates to the term reproduction setup as the number of loudspeakers and their spatial position.

In OBA there can still be conventional parts of the audio mix like an audio-bed[9] with a certain amount of channels, but in addition to that, single so called audio-objects[10] can be transmitted. These audio objects are not mixed or mapped to a specific channel of the reproduction setup, they are channel-independent audio files enriched with metadata and can be changed by e.g. spatial position or gain in the boundaries that the producer of the audio material has set in post-production. OBA can be played back via any device with any number of speakers and in addition offers the possibility of manipulating the audio objects. This also makes it very accessible and easy to use with the increasing amount of reproduction devices that exist nowadays (e.g. TVs, soundbars, home theatre systems, smartphones, tablets, laptops, etc.). A necessary requirement would of course be a viable infrastructure, such as an end device with implemented renderer software. Figure 4 below roughly illustrates the differences in the production workflow between CBA and OBA.

---

[9] The term audio-bed usually refers to the background sounds of an audio mix like e.g. ambient noise or music.

[10] Audio-objects are defined as audio combined with metadata like position information or gain interactivity information.

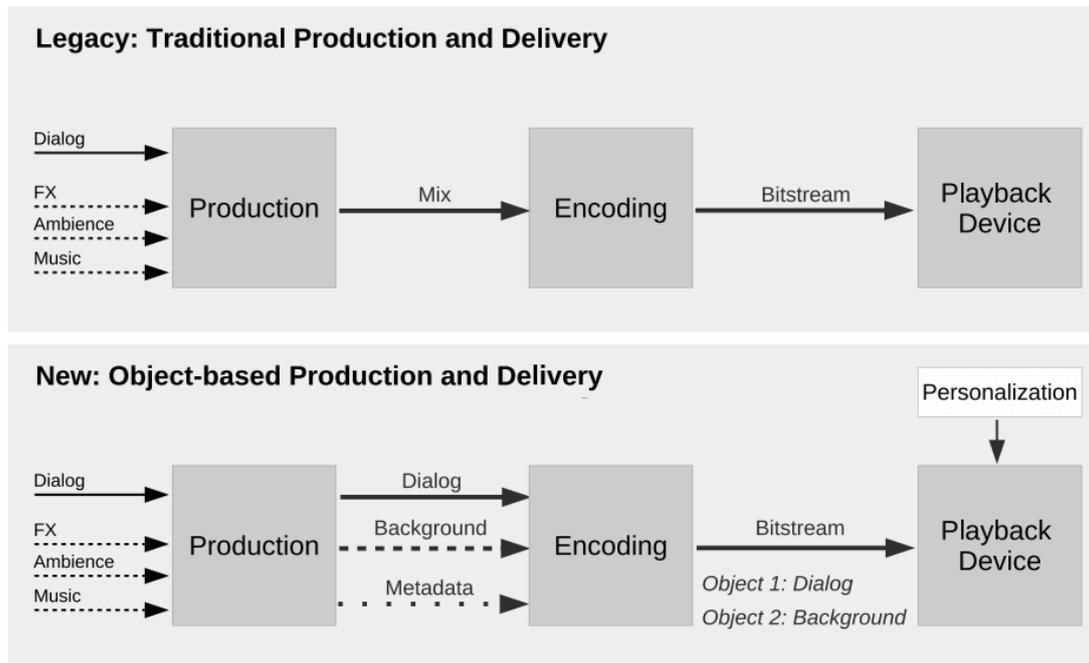

Figure 4: Differences between the production workflow of CBA and OBA reproduced from [43, p.4] with kind permission of the authors.

When talking about OBA, one often stumbles upon the term of Next Generation Audio (NGA). This term sums up audio systems that combine new audio technologies like OBA, proprietary codecs, encoding- and decoding-systems or a Dynamic Range Control (DRC). So OBA is a part of NGA. Some examples of current NGA systems with OBA support are e.g. MPEG-H Audio, Dolby AC-4 and DTS-UHD [10].

### 2.3.2. Possibilities of OBA with focus on accessibility

As shown in chapter 2.1, an important part of today's society suffers from some kind of visual or auditory impairment. Especially the growing amount of elderly people above 65 makes up an increasing part of the audience. It is important for these groups to be able to inform themselves and take part in culture. Broadcasting and streaming still are necessary components of participation in social events.
But modern broadcasting and streaming services often still use technology that is hardly capable of fulfilling the needs of a diverse and ageing audience. NGA-systems such as MPEG-H, that support OBA, offer a huge variety of possibilities to make audiovisual content more accessible and hence more suitable for increasingly heterogeneous viewers. Many broadcasting stations have their own editorial staff for viewer questions. These editorial offices often receive critique on speech intelligibility [18, 9]. Two separately carried out surveys from the year 2010 show that more than half of the participants (60% of 20,000 viewers) had difficulties with following spoken words on television [3].

The following table gives a quick overview of some of the most common accessibility barriers that are encountered with increasing age:

| Age-related factors | Resulting problems | Additional problems | Solutions |
|---|---|---|---|
| -Hearing loss | -Worse understanding of spoken language and non-speech sounds due to their low volume | -Audio level of speech and non-speech noise is too low | -Dialogue Enhancement |
| -Sight loss | -Disability of following the plot | -Consuming content in a noisy environment worsens understanding | -Audio Description or written subtitles |
| -Cognition loss | -Worse understanding of speech due to its complexity | -Artistic demands of producers limit speech intelligibility | -Simplified language |

Table 2: Common accessibility barriers that arise with age based on [43].

As the average age of the population increases, society is facing new challenges in a broad range of areas. First and foremost, the visual acuity decreases [31]. At the same time there is a decline in cognitive processing to be noted [28]. Due to presbycusis, the understanding of spoken language worsens. This results in a lower ability of following the plot of audiovisual content [11].

In addition to these age-related factors, there are other problems that make it difficult for older people in particular to follow audiovisual content on TV and streaming. There often is a significant difference between the artistic demands of producers or audio-mixers and the need to create content with well intelligible dialogue. This often results in a too low audio level of speech and non-speech noise in comparison to background music, sound effects and ambient noise. Sometimes also the listening environment can cause a problem. If e. g. the noise level in the surrounding environment is too high or the signal is reproduced via devices with poor playback quality, spectral masking can lower the intelligibility of speech as well. Lastly, listening to a program with foreign language normally requires more concentration and the listener would benefit from a higher SNR between the dialogue and background sounds.

If some of these factors are combined, it can get very hard to understand the spoken dialogue in the audio track of e.g. a film up to the point where the plot can no longer be followed properly. With OBA it is possible to offer additional audio tracks with e.g. simplified language to make it easier to follow the dialogue, audio description for helping visually impaired people hear what is going on in the picture or DE for a personifiable ratio between FG and BG loudness.

With NGA audio systems like MPEG-H it is not only possible to offer more different versions of the audio material for a bigger variety of viewers, it can also offer the consumer the possibility of choosing presets for different listening-environments or personalizing the audio mix to their own needs. For example, with the right preparation in advance, it is possible to give the end user the freedom of adjusting the speech volume to individual preferences and lower the loudness level of background music and sound effects [43, 56]. As one can easily see from the previous examples, OBA offers older people in particular the chance to stay in touch with cultural events via TV or streaming. In the end, this participation contributes to a feeling of well-being and independence.

### 2.3.3. Use cases and methods of enhancing speech intelligibility in OBA

When looking at the use cases of enhancing the intelligibility of spoken words on television and streaming, there are basically two important use cases to differentiate:

1. Broadcast content with accessible stems

2. Broadcast content without stems and only the full mix available

The first one mainly comprises of new productions, but also archive material. For the second occasion which is usually archive material, the project files with all necessary data no longer exist for all productions. Due to e.g. limited storage space and terminated contracts between production companies, in many cases only the complete audio mixes are available. The individual stems which would be important for preparing the material for speech intelligibility, are often no longer existing or not available. In addition to that, new productions are currently only produced object-based in exceptional cases because most broadcasters still lack technical know-how or the necessary infrastructure. This chapter will show a possible reprocessing workflow of already produced broadcast material with unavailable stems with regard to speech intelligibility.

According to the European Broadcasting Union (EBU), "hybrid approaches where channel-based mixes and objects are used simultaneously (...) might become one of the most often used scenarios" [10, p. 8]. Figure 5 shows a basic example of how such a hybrid approach of an NGA production could look like in the future.

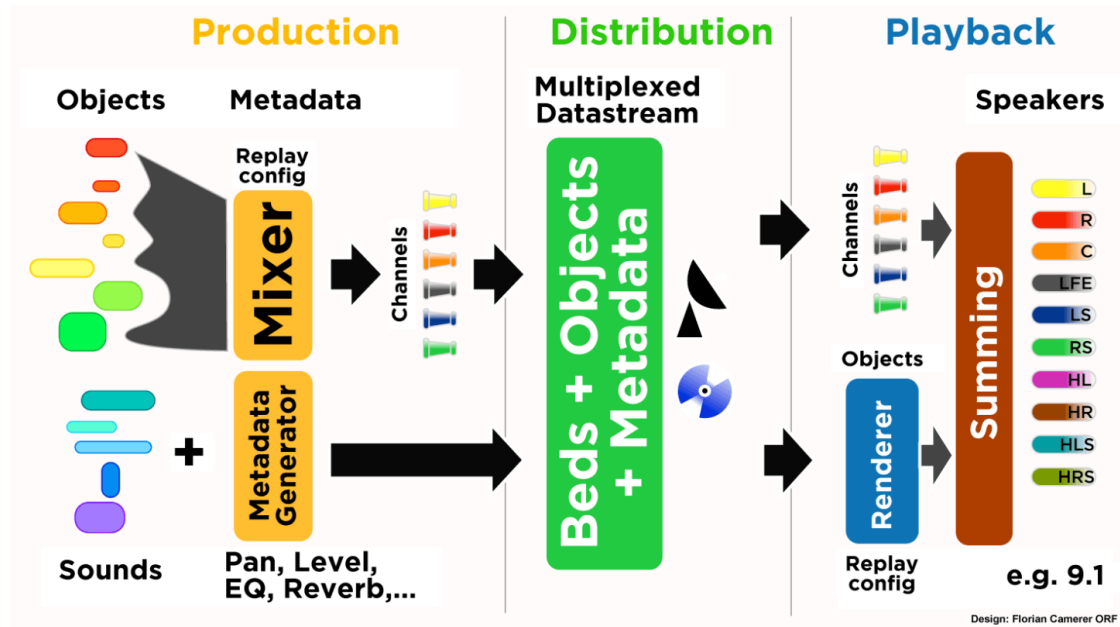

Figure 5: Basic principle of a hybrid NGA approach according to [10, p. 8], reproduced with kind permission of the European Broadcasting Union (EBU).

This approach basically divides into three phases:

1. Production phase (before encoding)
2. Distribution phase (after encoding)
3. Playback phase (after decoding)

When taking a look at how much archive material is streamed via internet and transmitted by broadcasting stations every day, it gets clear that this also is a very important part of the audio visual content, viewed by millions of people every day. As already stated above, the single stems are often not available any more. Instead, only a full mix is accessible. As a result, a re-adjustment of the balance of FG and BG in post-production is not possible any more, because the full mix contains all parts like speech, music or sound effects mixed into one audio file. In order to rehash the dialogue for DE, it is necessary to separate FG from BG. This can be achieved by using DS algorithms. This will be briefly explained below.

1. In the production phase, the full audio mix can be split up into two components by a DS algorithm. The FG then consists of an estimation of the original speech signal, whereas the BG consists of an estimation of all other non-speech sounds. The BG component could then be treated as a regular audio bed consisting of audio-objects with a static spatial position. This is then very similar to conventional CBA. Furthermore, the BG loudness could be manually re-adjusted or edited with some sort of automated processes like e.g. a side-chain compression. The estimated FG could be treated as a dynamic or static audio object as well. The mixing-engineer could add advanced user interactivity such as position or loudness interactivity metadata. In this phase he can also define the ranges in which the end user can personalize these parameters or create presets e.g. for different listening environments. Further control data like overall loudness parameters could also be added to the metadata.

2. In the distribution phase, all audio is then encoded with its corresponding metadata into a bitstream and will be further processed during the broadcast transmission.

3. In the playback phase, the audio stream will be decoded by software inside the playback device. According to the previously defined metadata FG and BG will be decoded and rendered to their spatial position inside the sound field created by the reproduction setup. Together with the respective production method and a playback device with the decoder software, this would allow the viewer to e.g. either select a preset or adjust the dialogue volume according to their preferences.

# 3. Related work

Since the topic of accessibility and speech intelligibility in TV has been researched in various directions over the past decades, this chapter offers a rough overview of a selection of the most important works in this field. Research is presented that on the one hand forms the basis of this thesis and on the other hand addresses a similar issue. Section 3.1 introduces an earlier conducted research on which this thesis is based. Section 3.2 will focus on the topic of accessibility and speech intelligibility for elderly and hard of hearing individuals. Section 3.3 is dedicated to the area of DE in OBA. Finally, section 3.4 concludes why the research carried out for this thesis has its own innovative value.

## 3.1. Test methodology A/ST

The listening test method used in this thesis was first introduced in a paper [52] by Torcoli et al. in 2017. The listening tests which were carried out as part of that work, build the foundation of this Bachelor thesis.
As also described in [53], their tests were conducted with younger test listeners of about 25 years median age without significant hearing-loss. The preferred LDs between FG and BG were measured. Two item categories were distinguished: material with available original FG and BG and items with FG and BG generated by DS. The listener satisfaction of the participants was also evaluated with regard to the possibility of personalizing the audio mix.
The preferred LDs ranged from 0 to 13 LU. It was also found that the listener satisfaction was significantly increased by the adjusted LDs.

## 3.2. DE for the elderly and hearing-impaired

DE for elderly people with age-related hearing loss also builds a central aspect of this thesis. Therefore in the following some important related work is presented, which introduces other approaches to this topic.

**Sound balances for elderly and hearing-impaired people**
Already in 1991 a study, conducted by the British Broadcasting Corporation (BBC) in collaboration with organizations for the hard of hearing, aimed to evaluate the preferred level differences of hearing-impaired viewers between speech and non-speech background in television audio [29]. Due to the fact that Loudness Units were not yet introduced at that time, the test results refer to the Decibel scale. The test was carried out with about 336 participants of an average age of just above 60 years. They were presented different versions of audio material with varying background loudnesses of -6dB, 0dB and +6dB relative to a default value. Afterwards they were presented questionnaires to evaluate their preferred versions of the test material. The results did not show any definite tendencies of the viewers regarding the preferred background level. The authors concluded that speech intelligibility is not strongly related to background level, but that the preferred SNR between speech and background is nevertheless related to the grade of hearing loss.

Another study from 2008, conducted in collaboration with the public broadcasting service of Japan (NHK), investigated the preferred audio balance between speech and background sound for nine younger test subjects of 20-25 years and nine elderly people of 60-73 years [27]. Monosyllabic tests for each individual ear were conducted to examine the deterioration of sound separation ability in noise due to ageing. The results revealed a connection between age, the decrease of perceiving important speech sounds and the level of background noise. A system was constructed that is able to indicate a sound balance between narration and background sound that is suitable for elderly listeners.

**An approach to DE by the Dolby Laboratories**

In a paper published in 2008 by the Dolby Laboratories, the characteristics of entertainment audio that are beneficial to the speech intelligibility of elderly people are highlighted in a literature review [34]. The outcome of this overview is that the only practical way for enhancing the speech intelligibility for elderly listeners is ensuring a good speech audibility. Furthermore, an algorithm is presented that identifies the channels in a surround sound audio signal that contain speech. In presence of speech the remaining channels are then attenuated by lowering frequencies that could interfere with the speech signal.

**ITC Clean Audio Project**

The Clean Audio Project introduced by the Independent Television Commission (ITC) ran from 2003-2006 and was an approach that based on the increasing spread of Dolby Digital 5.1 audio broadcasting. Among other features, the technology includes an application very similar to the previously presented research by the Dolby Laboratories. The tests carried out were presented in [41]. The 41 test subjects of 30 to 75 years of age with varying hearing levels were presented different versions of Dolby Digital 5.1 audio material with varying sound levels of the left and right speakers. The centre channel which contained the speech signal kept the same reference level. Afterwards the test subjects were asked to rate the overall sound quality, the enjoyment of the audio material and the clarity of the dialogue. The study observed that hearing-impaired people directly benefited from lowering the level of the side and surround speakers and that the overall rated sound quality and enjoyment was connected to the clarity of the dialogue.

**DTV4ALL and HBB4ALL**

DTV4ALL (Digital Television for All) was a project funded by the European Commission which was aimed to identify and push forward broadcast technologies and services that enhanced accessibility in television. Speech enhancement was also one of these topics. However the tests carried out showed the difficulties and problems with enhancing the intelligibility of dialogue [3]. The currently ongoing follow-up project of DTV4ALL is HBB4ALL (Hybrid Broadcast Television for All) [4]. It looks at how HbbTV, the European standard for broadcasting and broadband multimedia services, can be used to improve access services such as subtitling or audio description. The research looks at access services under laboratory conditions with end users as test participants. It also addresses the area of DE by looking at processing techniques to produce improved speech intelligibility. The advantages of personifiable audio mixes are evaluated as well [35].

**Further related work**

Other work has been published in the field of DE for addressing the raising number of people with hearing-loss in the society. For example, a 2016 released research and development white paper by the BBC that summarizes the history of the clean audio approach, states the difficulties that previous work came across and also looks at the progress that has been made in recent years [4]. A further work from 2019 reviews previously conducted research about accessibility in object-based audio and translates the stated findings into practical broadcast accessibility technology [56].

Finally, some production guidelines and recommendations have been published by various institutions and researchers to help broadcasters improve their content for speech intelligibility and accessibility. Some works include, for example, an International Telecommunication Union (ITU) report on how to make TV accessible [22], two diploma theses on speech intelligibility in television [18, 11] and a production guideline of the German broadcasting corporations ARD and ZDF on recommendations for improving speech intelligibility in television [9].

### 3.3. Dialogue enhancement within object-based audio

Since this thesis is dedicated to the central topic of dialogue enhancement in object-based audio, some important research and projects that also cover this topic will be highlighted in the following.

**MPEG-H**
The MPEG-H 3D Audio (MPEG-H) standard is a NGA system that supports dynamic and static audio objects as well as 3D audio formats such as Higher Order Ambisonics. In this system it is possible to personalize the audio mix by manipulating the single components of the sound like e.g. adjusting the background volume relative to the dialogue level to suit the individual needs of a diverse audience [30, 16].
A subsequent paper published in 2019 presented the implementation of DE in MPEG-H and put special focus on DS methods to use in case of FG and BG not being available separately [37]. The benefits for the users are investigated using an Adjustment / Satisfaction Test (A/ST). 14 participants of a median age of 25 years were shown seven different test items with speech in FG and varying types of BG such as ambient music or a cheering crowd. Again, just like in [52] two different ways of DE were used. Enhancement by using the originally available audio components and DE through BSS. The possibility of personalizing the audio mix was extensively used. Slightly lower LDs for DS items were preferred as for the test items consisting of originally available components. The authors conclude that "the benefits of object-based audio, as they are used in modern broadcasting systems, can also be used when broadcasting legacy content that was not produced in an object-oriented way by using current source separation technology" [37, p. 517].
A further publication of the year 2015 presented another object-based approach that could be used to personalize the audio mix within MPEG-H and hence support DE for hard of hearing viewers [40].

**Dolby AC-4**
Another well-known object-based audio codec is AC-4 from Dolby. According to their 2016 published paper their product is "designed to address the current and future needs of video and audio entertainment services, including broadcast and Internet streaming" [26, p. 1]. In addition, the audio system also offers the personalization of the audio mix and a DE technology with different modes of operation for variable bitrates.

**SAOC-DE**
A new specification was introduced in 2014 called SAOC-DE (Spatial Audio Object Coding for Dialogue Enhancement) which extended the already existing standard SAOC (MPEG-D Spatial Audio Object Coding), an audio coding algorithm which allows highly efficient storage and transport of individual audio objects, while offering the possibility of adjusting the mix based on ones personal taste [36]. Two independent listening tests were conducted to prove the functionality of the DE.

The first test comprises of a speech intelligibility test [13]. Ten participants with age-related hearing loss of a mean age of 73.3 years performed the Oldenburg sentence test (OLSA), whereas ten normal hearing listeners of about 23.3 years mean age functioned as a reference group. Sentences had to be understood in two different conditions of noise. Noise similar to the human voice and broadband noise. Starting from a default value the DE algorithm was used to attenuate the background noise at two levels: by 6 dB and 12 dB. For both test items and both attenuation levels the DE was able to increase speech intelligibility significantly for the elderly listeners and the reference group.

The second listening test was a Multi-Stimulus Test with Hidden Reference and Anchor (MUSHRA) [36]. The test items comprised of five different realistically conditions of speech and background noise like e.g. a commentator talking over an stadium ambience. Ten experienced participants performed the test. The DE was tested on speech amplification levels of +6 dB and + 12 dB. The results of the tests were all within the MUSHRA rating unit "Good".

**A&E Audio and Narrative Importance**
Another promising object-based approach on improving the TV audio mix for elderly and hearing-impaired people is the Narrative Importance Metadata [57]. In addition to enhanced dialogue and a BG attenuation, audio objects are enriched with metadata about the narrative importance of the audio. So, how important non-speech sounds are for the narrative of the plot. The idea behind this is to either offer an optional audio mix with amplified narrative important audio or to give the user control over the volume balance between the BG and the narrative important audio objects and dialogue. A&E (Casualty Accessible and Enhanced) Audio was a project in which this technology was first publicly tested [58]. In collaboration with the BBC an episode of a series was available online which could be changed seamlessly from the full mix into an accessible mix with narrative important enhanced audio by dragging one simple slider. Afterwards the viewers could complete an online survey. 299 people took the survey, 47% of whom suffered from any form of hearing loss. 73% of all participants reported that they were able to follow the content more easily or enjoyed the program to a greater extent due to the personalization option.

**Further related work**
Further work concentrates on specific DE methods such as dynamic BG ducking for speech intelligibility improvement, which could also be implemented in object-based audio [50]. In this publication again highly personal preferences in terms of the preferred LD are observed which emphasize the importance of personifiable object-based audio.

Other research has e.g. covered the challenges and requirements of dialogue control and the design of a corresponding user interaction interface [24] and e.g. the EBU has published a report on the need for broadcasters to implement NGA production workflows, not at least to encourage the issue of accessibility [10].

## 3.4. Conclusions for this thesis

As can be seen from the previous literature review, there has been done diverse research in the field of DE for broadcast audio. Hearing impaired and elderly people are also often in focus. For this thesis, the preferences of listeners aged above 65 in terms of their preferred LDs for TV programs have been evaluated. By using the A/ST as a test method, it was possible to give the participants a similarly free personalization experience as it would be possible with object-based audio. The two cases of originally available components and components created by DS were also compared. To the best of our knowledge, this is the first work of its kind.

# 4. Listening test

This chapter presents details about the conducted listening test.

## 4.1. Scope of the experiment

A listening test is being conducted, followed by a questionnaire. The listening test is firstly intended to evaluate if the possibility of personalizing the LD between FG and BG improves the overall listener satisfaction and if it makes it easier to follow the speech. As target user group, people aged 65 or above are considered. Secondly it evaluates if there are differences between stem-based and DS-based DE noticeable for elderly viewers and if there are preferences in terms of LD. The purpose of the questionnaire is to collect information to help interpreting and evaluating the results.

## 4.2. Test methodology

DE is based on the increase in the LD between FG and BG. Since preferences for LDs are highly dependent on individual and situational factors, it can be assumed that a standard preset for DE is not sufficient to cover the needs of a large number of people. In order to measure these inter-individual preferences, a test method is needed in which one can choose the own preferred LD. Furthermore, with OBA it is possible to let the viewer interact with the program, which also speaks for a test method that includes the personalization of the LD. In addition, the participants are selected to be 65 years of age or older. Elderly people tend to have less experience with the usage of technology and due to their age motor limitations are possible. Hence, it is also important to have a test procedure that is both simple to use and easy to understand. All these factors led to the decision to use the A/ST[53] as a test method. It is a multi-stimulus test for the subjective evaluation of user-adjustable systems based on the visual programming language for music and multimedia Max/MSP[1], developed by Cycling '74. A questionnaire is being used to gain more information about the test participants' perception of the test and to improve the test procedure itself for future use.

The A/ST contains multiple items, see chapter 4.4. For each item it is possible to change an adjustment parameter $p_a$. $p_a$ changes the LD between FG and BG of an item and can be varied from a default value into positive or negative direction, which means increasing or decreasing the BG loudness. This results in a variable loudness of the speech. The test participants can adjust $p_a$ by using a rotary knob. Per item, the they can rate their overall satisfaction of the personalized version $V_p$ compared to the default version $V_d$ by changing a second parameter for the assessment of the listener satisfaction $p_s$. $p_s$ changes values on a provided label scale reaching from "Much worse", "Worse" and "Slightly worse" over "The same as" up to "Slightly better", "Better" and "Much better". $p_s$ can also be changed by turning the rotary knob.

The A/ST basically comprises three phases:

1. Explanation and training phase

2. Adjustment phase

3. Satisfaction assessment phase

In order to be able to assign the results of the tests reliably to the corresponding test participants, the Participant-Identifier (P-ID) is entered before the test. Afterwards, the test participant sets the overall volume for the test to a similar level as he would watch TV at home. This volume may not be changed for the duration of the test.

1. In the explanation and practice phase, the test procedure, scenario, and how to operate the test are explained. It is very important to make sure the participants understand everything well to be able to produce reliable results. The first of all items is a training item. The values of $p_a$ and $p_s$ for this item are not taken into account in the final results.

2. In the adjustment phase, the participant first listens to $V_d$ and has to adjust $p_a$ to a value that makes speech easy to comprehend while maintaining a still enjoyable BG which results in $V_p$. If, due to the DS, the audio quality worsens, the participants are asked to find a good balance between speech intelligibility and overall audio quality. While adjusting $p_a$, the participants can A-B-compare $V_d$ and $V_p$. This helps the listener in understanding the adjustment being made. As the adjustments very smooth and gradual, it can be frustrating not having a reference point and the listener might ask himself if something is actually changing. Once the participants are satisfied with $V_p$, they have to push the knob in order to confirm their adjustments.

Figure 6 shows the Graphical User Interface (GUI) presented during the adjustment phase. The test was conducted in German. Therefore all texts and the provided rating scale presented in the GUI were translated into German. Screenshots of the German version of the A/ST can be seen in appendix A.

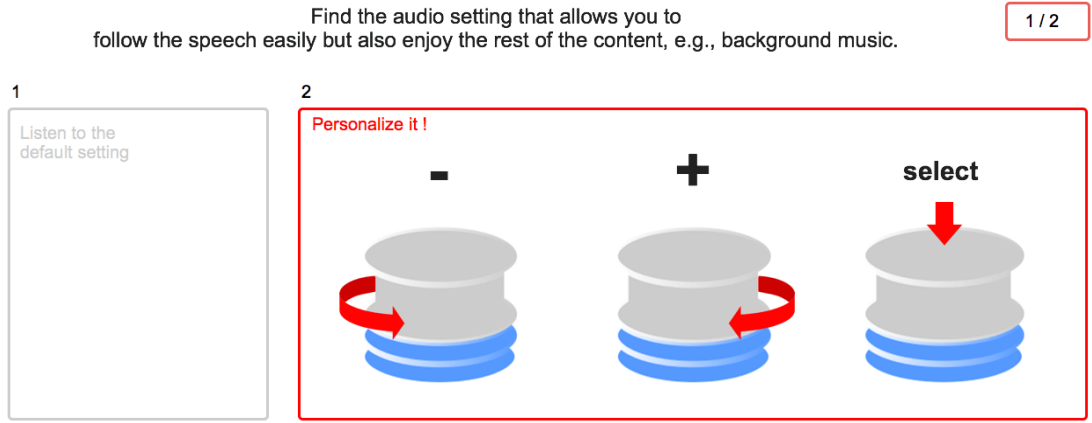

Figure 6: GUI presented during the adjustment phase of the A/ST [53].

3. In the satisfaction assessment phase, the participants are asked to rate $V_p$ against $V_d$ on the already above mentioned satisfaction scale. A functionality for A-B-comparison between the two versions is implemented in this phase as well. If the participant has selected a rating value, he can again confirm his selection by pressing the knob. In figure 7 the GUI for the satisfaction assessment phase is presented. The rating is being done in terms of how easily the participant can follow the speech, how much they enjoy the BG and what their overall satisfaction is. The results of this phase are also valuable for sorting out invalid data. If a $V_p$ is rated below the label "The same as", it indicates that either the participant got something wrong in the explanation and training phase or did accidentally personalize an item wrong in the adjustment phase, because the given task in this phase is to adjust and confirm $V_p$ only, if it is better than $V_d$ in terms of following the speech while maintaining a still enjoyable BG.

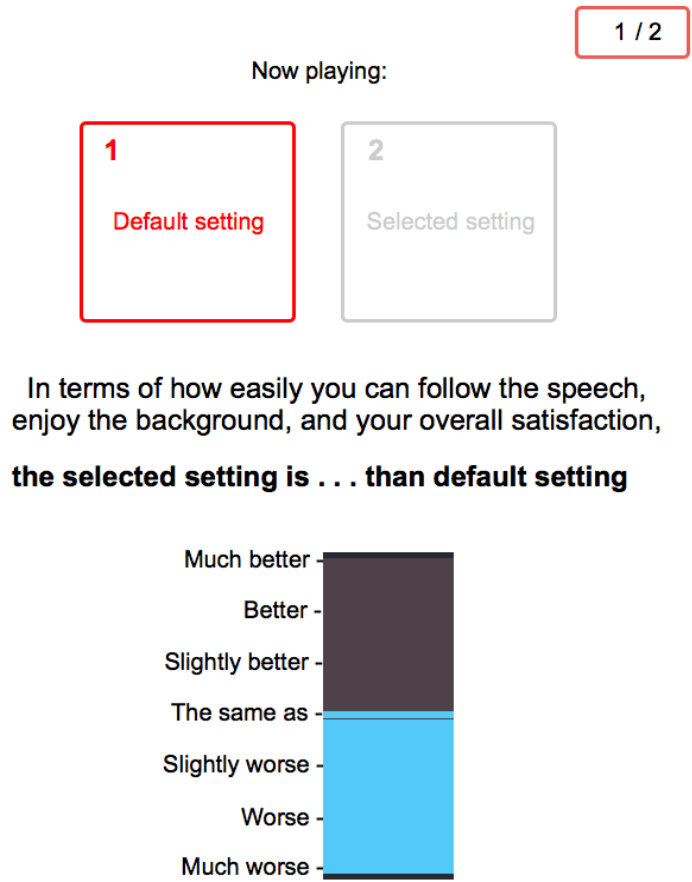

Figure 7: GUI presented during the satisfaction assessment phase of the A/ST [53].

For each item given, the participants work through phases 2 and 3 (adjustment phase and satisfaction assessment phase) as described above. A display in the upper right-hand corner of the GUI helps the participants to orientate themselves and to know which item they are at and how many more are to follow. The order of the items is not being randomized and stays the same for each participant. The A/ST outputs two results per participant and per item: a value for the preferred relative LD between FG and BG and a numeric value for the satisfaction rating which can be assigned to the corresponding label of the rating scale.

After the participants have completed the A/ST, a questionnaire is performed. The original document in German can be seen in appendix B. Each questionnaire sheet contains a cell for entering the P-ID. Seven questions are asked. The first one is numbered zero to indicate that this question has to be answered before the pure-tone audiometry. All following questions (1-6) are to be answered after having completed the A/ST.

Question number zero asks for a self-assessment of the own hearing. The participants can select the answers "excellent", "good", "average", "moderate" and " poor". This question aims to discover possible differences between the objective measurement and the subjective perception of the own hearing abilities, since some sources state that elderly people tend to ignore their proceeding hearing loss [48]. Question one asks: "How good was the user interface to understand and use?". This question provides valuable information indicating whether the complexity level of the listening test procedure was appropriate for elderly participants and whether further improvements should be made for future experiments of this type. Question number two asks if the test participant liked the possibility of personalizing the audio mix. This question evaluates the listener satisfaction more accurately and, for example, to make reliable recommendations to broadcasters as to whether older viewers would benefit from the possibility of personalizing the audio mix. Question number three asks: "What do you think was the purpose of this test?". Answers for this question can indicate how well the participants have understood the test scenario and can also help to double-check the validity of results of particular test participants. Question four asks what factors the participants based their personalization on. This question aims to find out how intensively the participants have already familiarized themselves with the subject matter and how differentiated they have applied their personalizations. As stated in different sources [18, 3], elderly viewers often have problems with understanding spoken words on television. To verify if this is also the case with the test group used in this experiment, question number five asks how often the participants have problems understanding spoken words on TV. The possible answers to select here, are: "Everyday", "At least once a week", "At least once a month" and "Never". Lastly there is room to add any sort of comments in question number six.

## 4.3. Feasibility study

For being able to evaluate if a test design is capable of producing the intended results and to check if e.g. the test items are working in a real-life situation, a feasibility study was conducted. The test procedure of the final listening test was developed incrementally. The main challenges that arose during this period are described in more detail in appendix C.

## 4.4. Items for the listening test

The conducted A/ST comprises sixteen different two-channel items being presented to the participants. They consist of the following categories:

| DE \ Prod. | **AR** | **WDR** |
|---|---|---|
| **OO** | $AR_{OO}$ (3) | $WDR_{OO}$ (5) |
| **DS** | $AR_{DS}$ (3) | $WDR_{DS}$ (5) |

Table 3: Item categorization from the perspectives of dialogue enhancement method (DE) and item production category (Prod.).

In table 3, two perspectives of categorizing the test material can be seen. The columns indicate the origin and production of the material, whereas the lines represent the type of DE that was used to create it. Artificially created (AR) stands for artificially created audio material that does not come from a broadcaster but was created in house for testing purposes. Westdeutscher Rundfunk (WDR) stands for original broadcast material provided by the broadcaster WDR. AR and WDR items are each existing in an Original Objects (OO) and a DS version. OO stands for the expression "original objects" and refers to test material where the original stems were mixed with different LD values, without using DS. DS stands for test material where FG and BG were generated by dialogue separation using a BSS algorithm developed by Fraunhofer IIS applied on the default mix. In the DS case the original stems are ignored and the case where the stems are not available is simulated. This is done so to be able to compare the listeners' choices on the same items for OO-enabled DE and for DS-enabled DE.

**$AR_{OO}$**: As indicated in table 3, there are three test items that where artificially created with separately existing FG and BG, the $AR_{OO}$ items.

**$AR_{DS}$**: To produce the corresponding three $AR_{DS}$ items, FG and BG of the $AR_{OO}$ where mixed together and afterwards separated with DS.

**$WDR_{OO}$**: As shown by the table above, there are five items of original broadcast material from the WDR where FG and BG were already existent as separate components, the $WDR_{OO}$ items.

**$WDR_{DS}$**: The corresponding five $WDR_{DS}$ items where again created by mixing together FG and BG of the $WDR_{OO}$ items and afterwards separating them by using dialogue separation. For the mixing of all OO items in order to produce full audio mixes for the dialogue separation, the digital audio workstation Nuendo [47] was used.

Adjusting the LD between FG and BG of an item is achieved by jumping constantly in time between many individual versions of the same item as the knob is turned. These versions differ only in their LD values. When the knob is turned in one direction, versions of the item are played with decreasing LDs. Turning the knob in the opposite direction increases the LD. The LD steps between the individual versions determine how seamless the personalization appears to the test participant and how far the LD can be changed in positive or negative direction. So, for each of the sixteen items presented in the test, there are many different versions that were created using MATLAB [49].

For all AR items, the adjustable LD range was 29.6 Loudness Units (LU). Reaching from a relative BG amplification of +9.6 LU from the default value to an attenuation of -20 LU. From $\Delta$+9.6 LU to $\Delta$0 LU the versions differentiated from each other in 0.2 LU steps and from $\Delta$-0.8 LU to $\Delta$-20 LU the versions differentiated in 0.8 LU steps from each other.

For all WDR items, the adjustable LD span was 52 LU. Reaching from a relative BG amplification of +12 LU from the default value to an attenuation of -40 LU. From $\Delta$+12 LU to $\Delta$-15 LU the versions differentiated from each other in 1 LU steps and from $\Delta$-16 LU to $\Delta$-40 LU the versions differentiated in 2 LU steps from one another.

The reason why AR and WDR items vary from each other in terms of the adjustable LD span and the LU steps of the item variations is that all AR items already appeared in an earlier conducted A/ST that was performed by younger test participants [53]. To be able to compare the results of that test with the A/ST conducted for this work, the same items where chosen. The larger LD adjustment span in the, specifically for this work created, WDR items was chosen for having a higher freedom of personalization and for evaluating if more extreme LDs would be selected. The LD steps for the test items can be seen in more detail in the appendix D.

In the following Table 4 the content of the items is displayed. Column one contains the numbers of the items according to their appearance in the listening test. Column two shows the labels for each items, as it is displayed in the graphs in chapter 5. The third column contains the used DE method for each item, whereas column four contains the corresponding production category. The last column to the far right shows information about what sort of audio material each item is made of. A key for the shortcuts used in this column is given below the table.

| Item number | Label in graphs | DE method | Production type | Content |
|---|---|---|---|---|
| Training 0 |  | OO | WDR | mVO, noise |
| 1 | WDR3 | OO | WDR | mVO, music |
| 2 | AR2 | OO | AR | mVO, noise |
| 3 | WDR1 | DS | WDR | mVO, noise |
| 4 | WDR4 | OO | WDR | mVO, music |
| 5 | AR3 | OO | AR | mVO, noise |
| 6 | AR2 | DS | AR | mVO, noise |
| 7 | WDR2 | DS | WDR | mVO, music |
| 8 | WDR3 | DS | WDR | mVO, music |
| 9 | WDR5 | OO | WDR | fVO, mVO, music |
| 10 | AR1 | DS | AR | fVO, noise |
| 11 | WDR4 | DS | WDR | mVO, music |
| 12 | WDR5 | DS | WDR | fVO, mVO, music |
| 13 | WDR2 | OO | WDR | mVO, music |
| 14 | AR1 | OO | AR | fVO, noise |
| 15 | WDR1 | OO | WDR | mVO, noise |
| 16 | AR3 | DS | AR | mVO, noise |

Table 4: Item categorization with information about the content.

Key for table 4:
**fVO** = female Voice Over
**mVO** = male Voice Over
**music** = music or music-like sound effects present in background
**noise** = noise like e.g. ambience, cheering crowd, applause, rain present in background

For each test, the order of the sixteen items followed the numbering of the items in table 4. Furthermore, the integrated loudness of all items and the versions they consist of was levelled according to ITU-R BS.1770-4 [21] to prevent sudden loudness jumps when switching from one test item to the next.

## 4.5. Test participants

According to the WHO and other sources, hearing beings to deteriorate significantly starting at the age of about 65. Likewise, about one third of all people above 65 are affected by a disabling hearing loss [7, 59]. In addition, presbycusis can be observed increasingly more often in people above this age. Hence, the age limit of 65 years is ideal for testing the effects of DE in OBA on elderly people.
Another precondition for the participants was that they would not use any form of hearing device. This excludes the complex interactions of hearing aids on the test procedure. Furthermore, many elderly people do not use hearing aids, even if they would benefit from them [17].
The test subjects for this experiment were recruited from the Fraunhofer IIS test listener database. In total, 12 test participants were invited and performed the test procedure. The results of one participant had to be discarded because the results indicated that the test subject did not understand the procedure properly. This results in a usable number of participants of 11 test listeners. For each participant, audiograms where measured right before the listening test. Further information on this topic will be given in chapters 4.6 and 5.1. The age span of all participants whose results could be used is 9 years, reaching from 66 to 75 years. The mean age for the test subjects therefore lies around 70 years.

## 4.6. Execution of the test

As already stated above, the whole test was conducted in German language with German test subjects. After the participants had arrived, they were familiarized with the test scenario. In this context, the participants were also asked question zero of the questionnaire, which refers to the self-assessment of the own hearing. After that, a pure-tone audiogram was measured for each participant, using a silent measurement booth and a specialized software, developed by Fraunhofer IIS. Each participant was assigned an P-ID, which made it possible to associate the test results with the respective test subject data afterwards. Subsequently, the test persons were brought into the listening test room, which was acoustically treated with diffusers and absorbers in order to achieve a balanced sound and to minimize room-related acoustic disturbances. RT60 measurements of the room revealed a factual average reverberation time of 0.21s between 50 Hz and 4kHz. The participants were then given an instructional sheet which explained the test procedure in more detail. The original document in German language can be seen in appendix E. While working on the training item, the experimenter was still in the room to answer any potential questions. The remaining 16 items were worked on by the test subjects alone. After having accomplished the listening test, the remaining six questions of the already mentioned questionnaire were performed. The sitting position of the test participants was chosen equidistant to the two loudspeakers. These were set up in a position of 0 degrees elevation and -/+ 30 degrees azimuth to the test listener at a distance of 1.6 meters. The loudspeaker type used was Dynaudio BM 12A. The monitor screen was mounted in a position of less than 0 degrees elevation below the angle of projection of the loudspeakers, for causing the least possible influence on the sound propagation. Computer mouse, keyboard and the rotary knob were easily accessible on a small table in front of the participants. Appendix F shows the listening test room and a participant while working on the test. The flowchart in figure 8 gives an overview of the interaction of the hardware and software components used in the test.

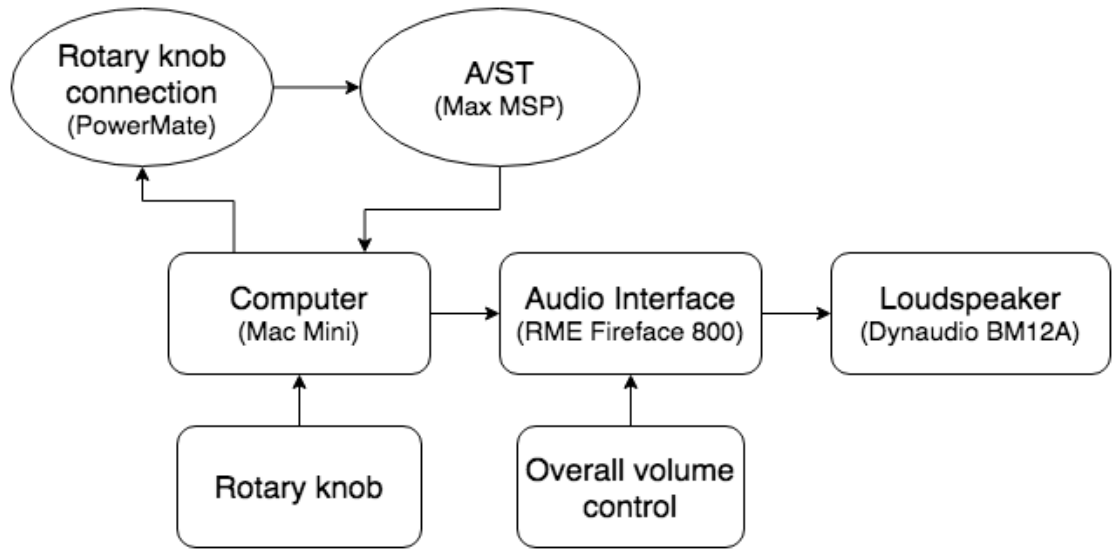

Figure 8: Soft- and hardware components used in the listening test.

The used computer was a Mac mini on which the Max patch for the A/ST was running. The rotary knob was connected via USB to the computer. A software called PowerMate established the connection between the motions of the rotary knob and the Max patch which played back the selected items. The audio was forwarded to the RME Fireface 800. Connected to this audio interface was a volume control knob, with witch it was possible to adjust the overall volume at the beginning of the listening test. As the final playback components, two Dynaudio BM 12A loudspeakers were used.

# 5. Results and discussion

Considering the close connection between discussion and interpretation of the results, the two sections will be combined in this chapter.

## 5.1. Audiogram measurements

As already mentioned under point 4.6, on each participant a pure-tone audiometry was performed. In order to be able to draw conclusions about the collected data and to get an overview of the participants' hearing, an average audiogram was calculated. Figure 9 shows an audiogram in which the frequencies from 125 Hz to 16 kHz are plotted on the x-axis and the hearing threshold is displayed in dBHL[11] on the y-axis.

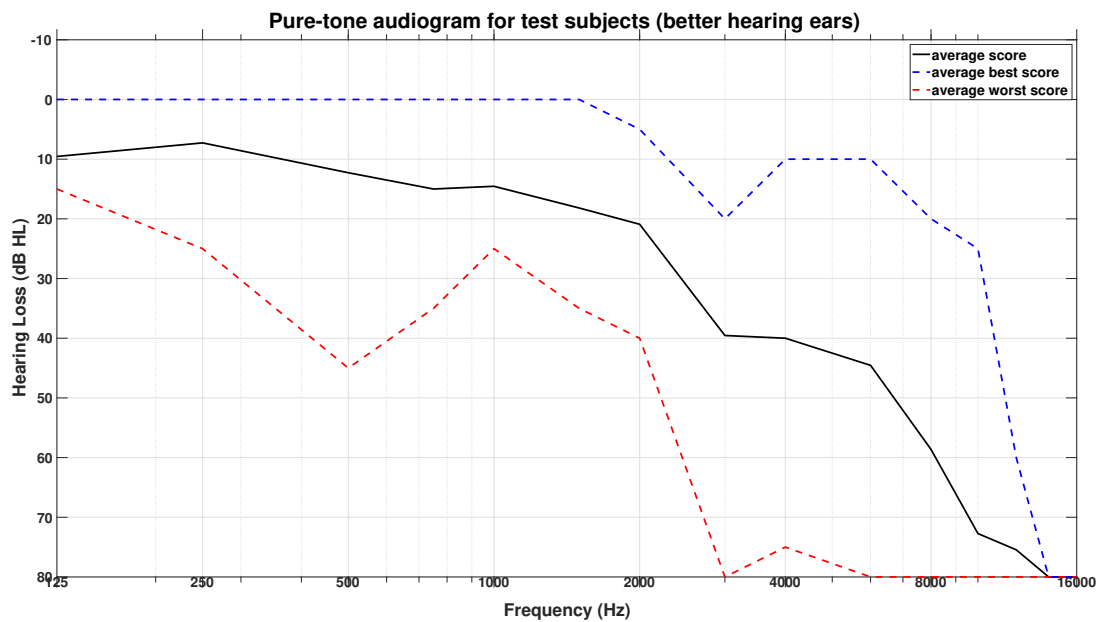

Figure 9: Average audiogram for all participants (n=11).

The presented data show the achieved values of the test persons' better hearing ears in a pure tone audiometry. The grey line indicates the mean audiogram for the better hearing ears of all 11 participants. The blue dashed line shows the average best values, whereas the red dashed line shows the average worst values of the better hearing ears of all participants. The graph shows that the test subjects have an average hearing without significant losses up to about 1.5 kHz. From this value on, hearing sensitivity decreases significantly in the higher frequencies. This is a typical pattern of age-related hearing loss.

---

[11]dB HL refers to the hearing threshold or hearing level. Hearing threshold is defined as the intensity at which the test subject can barely perceive a sound. Measured at different frequencies, the result is a hearing level curve (= Pure-tone audiogram).

## 5.2. Personalization of the LD in the A/ST

The following figure 10 displays the results of the A/ST, regarding personalization of the LD.

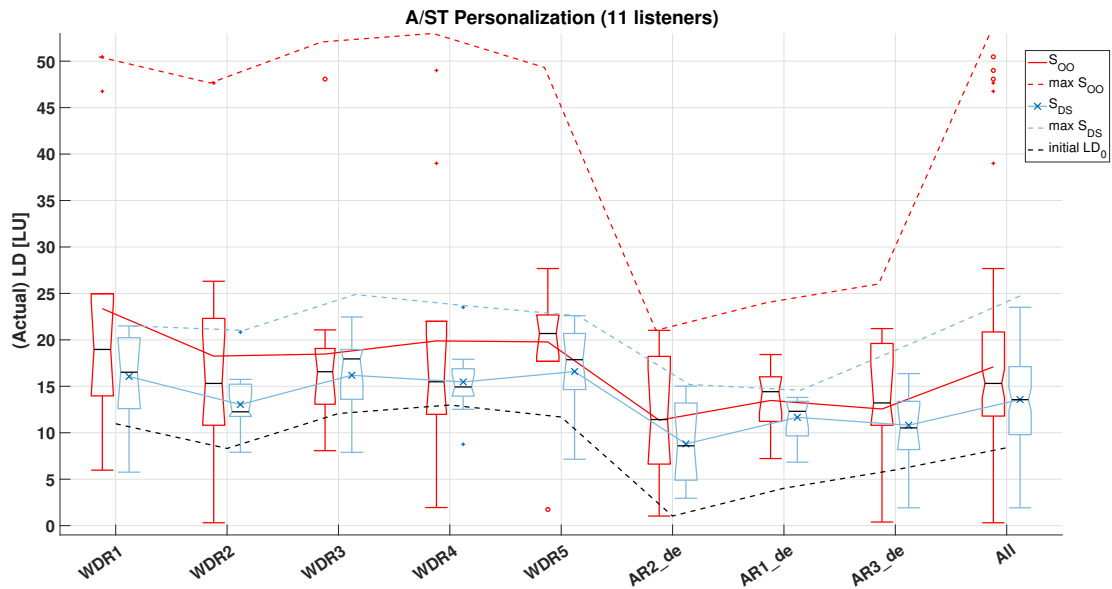

Figure 10: Box plots regarding the personalization preferences in the A/ST.

The diagram shows box plots[12] which present preferred LDs between FG and BG for each presented item. The x-axis shows the two item production categories WDR and AR, already introduced in chapter 4.4. The red box plots represent the data for the items of the DE category OO, whereas the blue box plots display the data for the items of the DE category DS, also explained in chapter 4.4. On the far right of the figure, there are two box plots labelled "All". These plots summarize the selected LDs for all items. The red solid line indicates the arithmetic mean values for the preferred LDs of the OO items. The red dashed line indicates the maximum personifiable LDs for the OO items, which is 40 LU relative to the default LD for the WDR items. In the same way, the blue solid line represents the arithmetic mean values for the preferred LDs of the DS items, whereas the blue dashed line indicates the maximum adjustable LDs for the DS items. Due to the fact that in DS there is a signal leakage of FG into BG and vice versa, the available maximum LD for the DS items is smaller than that for the OO items. The arithmetic mean values for the DS items are in addition marked with a blue X. The black dashed line indicates the default or initial LD from which the subjects could increase or decrease the BG loudness corresponding to the LD used in the real-world broadcasting of the WDR items. For a more detailed look at the collected raw data of the personalization in the A/ST there is a figure in appendix G, which shows all collected data points in relative LD values.

---

[12] A box plot is a compact way of visualizing the distribution of data points. Here the box is depicted vertically and hourglass-shaped. Its lower end corresponds to the first quartile Q1, the central bar (in this here work coloured in black) corresponds to the median, and the upper end corresponds to the third quartile Q3. Hence, the height of the box corresponds to the Interquartile Range IQR = Q3 - Q1. Vertical lines (often referred to as whiskers) extend from the box indicating the variability outside the upper and lower quartiles; they are concluded with horizontal bars positioned at the maximum or minimum point within 1.5 Interquartile Range (IQR). Points outside the whiskers range are displayed with a cross if they are between 1.5 and 3 times the IQR and with a circle if they are outside 3 times the IQR [53].

### 5.2.1. Differences and similarities between OO and DS in terms of preferred LD

On closer examination of the median values of the preferred LDs, it is noticeable that although they are close together for each item, they still differ from each other. For the OO items a larger median LD of 15.3 LU can be observed than for the DS items (13.5 LU). Hence, there is only a slight gap of 1.8 LU between the median preferred LD for the OO items and the median preferred LD for the DS items. When looking at the overall IQRs, there is a higher variability in the chosen LDs when working with the OO items (11.8-20.8 LU) than when working with the DS items (9.8-17.1 LU). Also, the IQR spans of OO and DS items vary by only 1.7 LU. Although it is not the purpose of this work to draw comparisons to similar studies, it is interesting to note that in an earlier conducted A/ST [53] with younger test listeners of 25 years median age, similar loudness gaps between OO and DS items were observed. However, the preferred LDs were significantly smaller (0-13 LU). It is also interesting to note that the LD values of about 7-10 LU or 10-15 LU, suggested by other sources [18, 51], were even exceeded by the older test participants of this experiment.

According to the instructional sheet, the participants were asked to select a LD level where speech is well intelligible, but at the same time the overall audio quality is also still enjoyable. Due to the working principle of DS algorithms, at the moment FG and BG do not sound as effectively and precisely separated as original objects. There still appear disturbing artefacts when trying to achieve high BG attenuation values in dialogue separated audio material. This could explain why the participants chose less extreme attenuation values when working with the DS items. One conclusion that could be drawn from the fact that the mean and median LDs of the single items are different but still very close together is that DS worked well for the test listeners over 65. The comparable overall median LDs suggest that the test subjects tried to achieve similar attenuation values when working with the DS version as when working with the OO version of the items. The two systems seem to offer very similar services for elderly listeners.

### 5.2.2. Preferred LDs vary greatly between participants

When looking at the variability inside and outside the IQRs for each item and also for the summarizing box plots on the far right of the graph, one can see that there is a huge variation across participants in terms of their preferred LD.
Every auditory system is individually different and affected by hearing loss to varying degrees. In addition, the personal taste in terms of the sound mixture also varies from person to person. Listening experience and possible hearing training are also decisive factors. All these aspects can explain a strongly individualized preferred LD and are an indicator that older test listeners above the age of 65 could benefit from the ability of personalizing the audio mix.

### 5.2.3. Preferred LDs vary greatly within a participant

As shown in 5.2.2, a high variability between participants can be noted. But also a great range in terms of the preferred LD within each test subject can be viewed, when taking a look at the figure provided in appendix G. As already stated above, the graphs in this figure do not refer to absolute LDs, but to BG attenuation values, relative to the default BG loudness. Nevertheless, from the graphs labelled "Listeners" it can be observed that the preferred LDs within the single test listeners vary greatly for both the OO and DS items. However, this does not mean that the individual participants have chosen their personalizations at random. The selected LD values within a participant have a wide distribution, but are consistent for OO and DS items.
One possible explanation is that different audio material also requires different attenuation values to make the language comprehensible. For example, a constantly low frequent atmosphere outside the human voice's frequency spectrum will need to be attenuated much less than e.g. a BG which contains many parts of the speech frequencies. Moreover, in this matter the influence of personal taste in terms of the sound mixture is certainly a particularly important factor.

### 5.2.4. Preferred LDs vary across items

Isolated from the item production and the DE category, there is a variation noticeable, in terms of the preferred LD. As can be seen from the variance of the results for different items, the preferred mean and median LDs vary from item to item and also the range of values varies.

These phenomena can be explained by the preferred LD being strongly dependent on the audio material. In addition, the individual hearing ability also plays a decisive role. Since a wide range of diverse material is broadcasted on television, this result supports the implementation of personifiable DE in existing workflows.

### 5.2.5. Overall preferred LDs of OO and DS vary from overall default LD

When comparing the black dashed line which indicates the default LD of the items, and the corresponding solid red and blue lines, a significant difference between the default LDs and the adjusted LDs becomes apparent. The mean default LD of all items is 8.4 LU, whereas the mean LD of the OO items is 15.3 LU and the mean LD of the DS items is 13.5 LU. This results in a difference of 6.9 LU between the mean default LD and the preferred mean LD for the OO items and a difference of 5.1 LU between the mean default LD and the preferred mean LD for the DS items. It can therefore be concluded that the test subjects not only accepted, but intensively used the possibility of personalizing the LD. Since the instruction was to change the sound mix to make the speech more comprehensible, it can further be concluded that both systems, DE by OO and DE by DS, were able to improve the clarity of the speech. It can also be pointed out that at least for the WDR items which consist of authentic broadcast material, the original LD between FG and BG is not sufficient for elderly viewers. The already stated observation that most of the adjusted LDs lie between 9.8 and 20.8 LU (IQRs of OO and DS items combined) indicates that elderly listeners above 65 could benefit from a large LD personalization range, presumably due to age-related hearing loss.

## 5.3. Satisfaction assessment in the A/ST

The following figure 11 presents the collected and processed data for the satisfaction assessment in the A/ST.

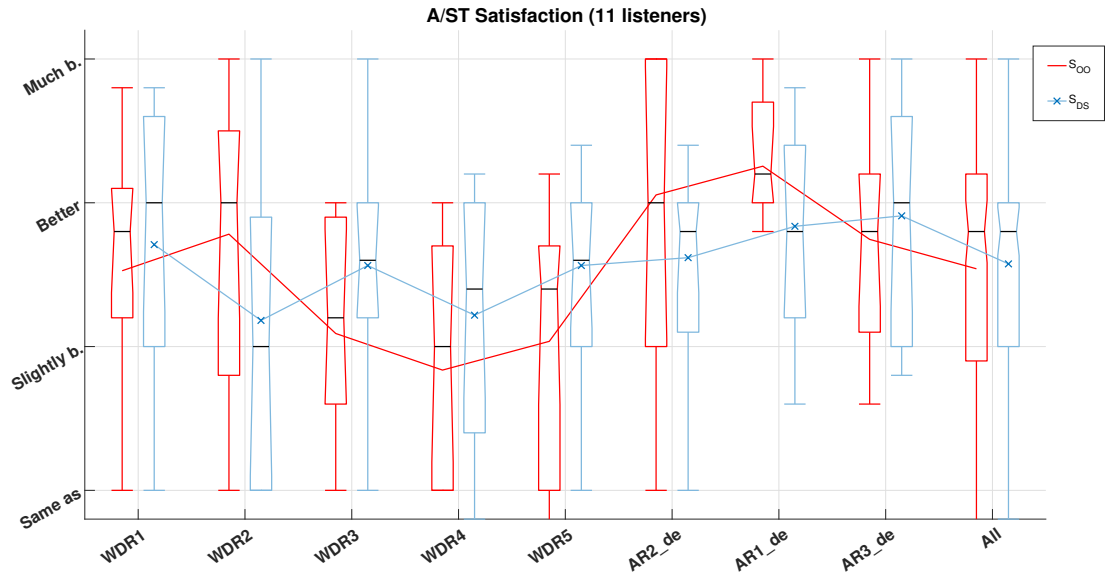

Figure 11: Box plots showing the collected data regarding the satisfaction assessment in the A/ST.

On the y-axis, the rating labels of the upper half of the satisfaction rating scale (chapter 4.2) are plotted. The x-axis shows the two item production categories WDR and AR, already introduced in chapter 4.4. The red box plots represent the data for the items of the DE category OO, whereas the blue box plots display the data for the items of the DE category DS. On the far right of the figure, there are two box plots labelled "All". These plots display the data for all items in a summarizing manner. The red solid line indicates the arithmetic mean satisfaction values of the OO items while the blue solid line stands for the arithmetic mean satisfaction values of the DS items. Again, for a more in-depth review of the raw data on the satisfaction assessment in the A/ST, there is a figure in appendix H which displays all collected data points.

### 5.3.1. Most median satisfaction values are located in the upper middle range

When looking at the median satisfaction values for OO items as well as for DS items, a tendency towards the rating level "better than (default value)" is identifiable.

The median satisfaction values of the overall box plots on the right hand side of the graph also illustrate this observation.

As the assessment should be based on language comprehensibility, background enjoyment and general satisfaction, this result indicates that the two DE systems are able to provide an improvement in terms of these factors.

### 5.3.2. Overall median satisfaction of OO and DS is identical

A look at the median satisfaction values induced by the two DE methods shows that they share the same value. Displayed on the right hand side of the figure, the numerical overall median values for OO and DS items are both 24 which is next to 25, the numerical value assigned to the rating "Better as (default value)".

On the one hand, this indicates that the test listeners were overall satisfied and that the two systems could provide an improvement in being able to follow speech beyond the default value. On the other hand, it shows that the test listeners were equally satisfied with the personalization results of both systems on average and that the increase in speech clarity with DE by DS worked just as well as the equivalent with OO.

### 5.3.3. For each item the median and mean satisfaction of DS and OO are closely related

Not only the overall median satisfaction of OO and DS items is similar, also the median and arithmetic mean values per each item are closely related to each other. Looking at the red and blue solid lines, it is evident that they follow each other and share a similar value range throughout all items.

This leads to the conclusion that the two DE systems follow each other in their mode of effect and are able to generate similar satisfaction values in the test persons even for different kinds of audio material.

## 5.4. Questionnaires

In the following section the results of the conducted questionnaires will be presented and discussed.

### 5.4.1. Self-assessment of own hearing

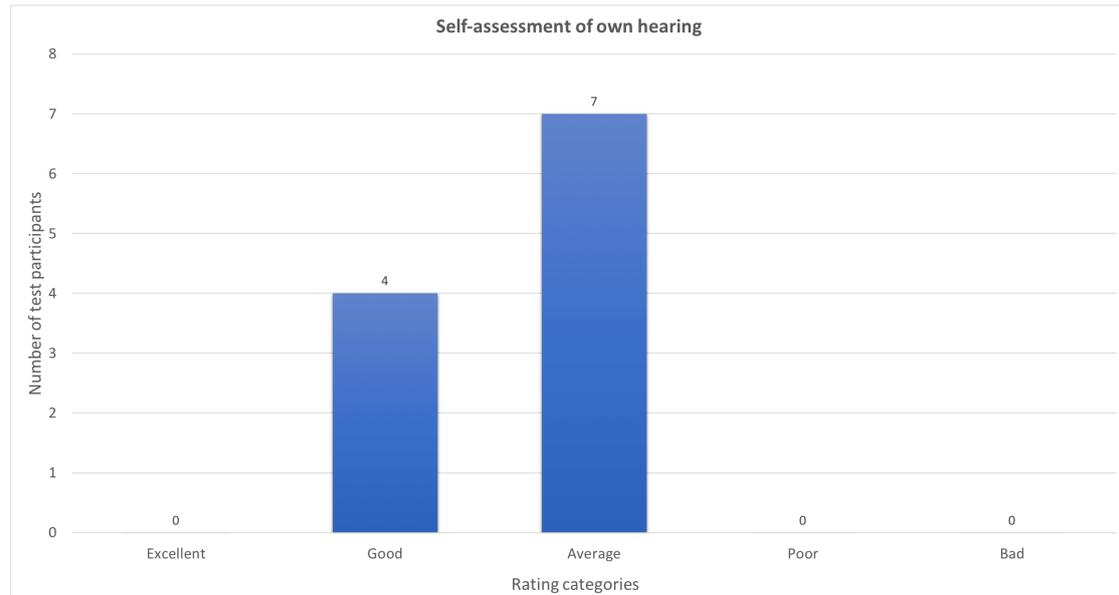

Figure 12: Self-assessment of own hearing ability of all 11 test participants.

Figure 12 contains the participants' answers for question number zero of the questionnaire: "How do you assess your own hearing ability?". This question was asked before the audiograms were measured, in order to obtain an unbiased answer. The five response options "Excellent", "Good", "Average", "Poor" and "Bad" are plotted on the x-axis. The y-axis shows the number of test subjects. Of eleven participants, more than half of them assessed their own hearing as average. Four participants rated their own hearing as good.
This result can be considered good, as the older participants assessed their hearing relatively realistically overall. When asked why many respondents rated their hearing as average, some responded that they considered their hearing performance to be average for their age group.

### 5.4.2. Problems understanding the language on TV

Figure 13 shows a bar chart with the results for the question "How often do you have problems understanding the language on TV?". On the x-axis the four response options "Every day", "At least once a week", "At least once a month" and "Never" are plotted. The y-axis again shows the number of test subjects.

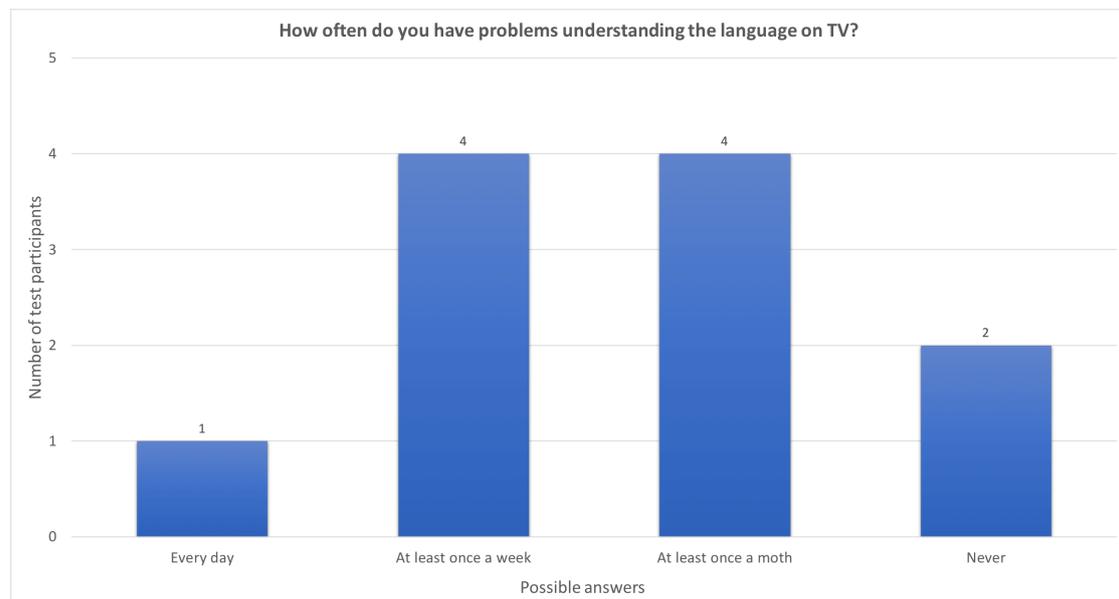

Figure 13: Answers for how often all 11 subjects have problems following the language on TV.

Disregarding the regularity, a majority of 9 participants (81.7%) reported that they have problems following the language on television. Two test participants declared to never have difficulties with speech intelligibility on TV.
When asked about their answers, one of the two people who answered "never", explained that he hardly ever had problems with following the speech on TV because he hardly ever watches television. Assuming that the audio mixes on television should be produced for the widest possible audience in order to exclude as few people as possible, and elderly people above 65 constitute an increasing part of the TV viewers [2], it underlines the need for implementing a DE functionality in existing broadcast production and transmission chains.

### 5.4.3. Personalization of the sound mix

Below another figure (14) is shown. It displays the answers of the test listeners to the question "Did you like the possibility of personalizing the sound mix? On the x-axis the possible responses "No" and "Yes" are given, while the y-axis shows the number of test subjects.

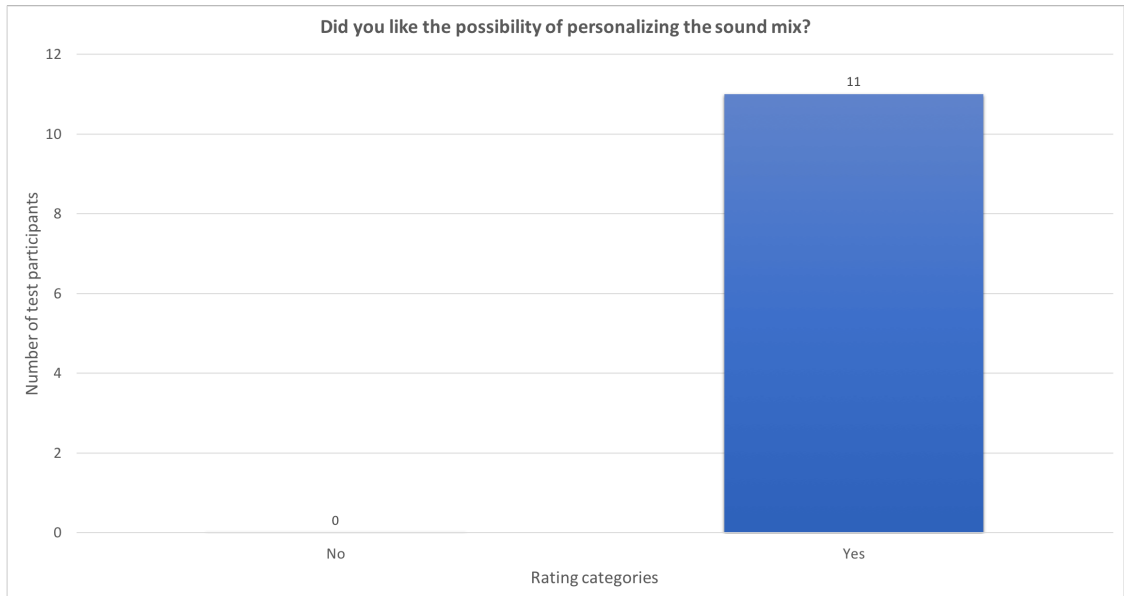

Figure 14: Answers of all 11 participants for the assessment if they liked the possibility of personalizing the audio mix in general.

Without exception, all eleven participants said that they enjoyed the possibility of controlling and changing the default audio mix to their own taste and needs. Some test subjects explicitly added that they would enjoy such a controlling possibility for their television at home. One participant mentioned in particular that the German TV crime series "Tatort" would benefit from this opportunity, as the dialogue here is often difficult to understand. This unambiguous result shows that elderly people above 65 could benefit from the DE enabled option of personalizing the audio mix.

# 6. Conclusion and outlook

First of all, this work took a look at the problems and challenges that elderly viewers currently face when watching television due to age-related hearing impairments. It was pointed out that elderly people above 65 make up a large and growing part of the world's population and that hearing loss will therefore also increase. It was also pointed out that age-related hearing loss can cause difficulties in following speech on TV, and that this is currently a major point of criticism for broadcasters. A promising approach to solve some of these problems was presented, which consists of a combination of object-based audio and dialogue separation. It was stated that by using audio objects, this technology is able to give the end user the possibility of personalizing the audio mix. Regarding the authoring of object-based audio, a brief introduction was given to the two main categories of broadcast audio material, consisting of material with existing audio components and material with only the full-mix available. It was shown how object-based audio in combination with dialogue enhancement technologies can make the language on TV easier to follow and can offer the possibility of personalizing the audio mix.

Next, the test procedure developed to evaluate the research question was explained in more detail. Two different kinds of dialogue enhanced test material were used for testing: On the one hand, items with already existing audio components and on the other hand, material with foreground and background created by dialogue separation. The test participants could then first personalize the loudness difference between speech and background sounds of the audio material and then assess how satisfied they were with their chosen personalization.

Looking at the overall results, it can be concluded that elderly people above the age of 65 like the possibility of personalizing the audio mix and also use it extensively to customize the audio mix according to their wishes, if offered. It can also be said that the possibility of customizing the audio mix provides an increase in satisfaction. Furthermore, the personalization through dialogue separation has been well accepted and could provide as much satisfaction as the personalization through originally available audio components. It turned out that the favoured loudness differences vary greatly from one individual to another. The tests also showed that the loudness difference of the used original broadcast material is not sufficient for elderly viewers. All these aspects recommend the implementation of dialogue enhancement in broadcast workflows.

In his white paper from 2011, Mike Armstrong reviewed literature concerning the area of audio processing and speech intelligibility. In this work, he concludes that "current audio processing techniques cannot significantly improve the intelligibility of speech in noise, if at all" [3, p. 6]. This work has shown that today's audio processing techniques are in fact capable of improving the listeners satisfaction and can make the language on TV easier to follow. It shows that great progress has already been made in this area and that more research has to be done in that field. For future research it would be insightful to present video material in addition to acoustic stimuli and to evaluate the influence of video on the preferred loudness difference and the listener satisfaction. Also, similar test scenarios under realistic conditions would be revealing. Conducting a test in a living room with audio presented via the built-in speakers of a TV could be one example. Furthermore, it would be fruitful to study the influence of hearing devices on the preferred loudness difference more closely. In the listening test conducted for this thesis, items with statically attenuated backgrounds were used. In future works it would be interesting to see the differences that a dynamically attenuated background would evoke. Finally, the technologies tested individually for this thesis will have to be tested in the context of a complete real-life broadcast workflow.

# References

[1]  Cycling '74. *CYCLING '74: TOOLS FOR SOUND, GRAPHICS AND INTERACTIVITY*. 2020. URL: https://cycling74.com/ (visited on 02/27/2020).

[2]  AGF and GfK. "Fernsehkonsum: Tägliche Sehdauer der Deutschen in Minuten nach Altersgruppen". In: *Statista* (2019). URL: https://de-statista-com.proxy02a.bis.uni-oldenburg.de/statistik/daten/studie/2913/umfrage/fernsehkonsum-der-deutschen-in-minuten-nach-altersgruppen/.

[3]  Mike Armstrong. *Audio Processing and Speech Intelligibility - Research Whitepaper WHP 190*. Tech. rep. April. BBC, 2011.

[4]  Mike Armstrong. *From Clean Audio to Object Based Broadcasting - Research & Development Whitepaper WHP 324*. Tech. rep. September. BBC, 2016.

[5]  Audionamix. *Audionamix - XTrax Stems 2*. 2019. URL: https://audionamix.com/technology/xtrax-stems/ (visited on 11/20/2019).

[6]  Audionamix Inc. *Audionamix - ADX SVC*. 2016. URL: https://audionamix.com/2016/11/28/adjust-dialogue-within-full-mix-adx-svc/ (visited on 11/20/2019).

[7]  Ellen B. Braaten. *Deafness and Hearing Loss*. 2018. DOI: 10.4135/9781483392271.n116. URL: https://www.who.int/news-room/fact-sheets/detail/deafness-and-hearing-loss (visited on 01/23/2020).

[8]  Thomas Brand and Birger Kollmeier. "Efficient adaptive procedures for threshold and concurrent slope estimates for psychophysics and speech intelligibility tests". In: *The Journal of the Acoustical Society of America* 111.6 (2002). DOI: 10.1121/1.1479152.

[9]  Michael Eberhardt et al. *Sprachverständlichkeit im Fernsehen Empfehlungen für Programm und Technik*. Tech. rep. Stuttgart: ARD; ZDF, 2014.

[10] EBU. *Why Broadcasters need an open, Codec-Independend Workflow for NGA Production Deployment - TR 045*. Tech. rep. January. Geneva: EBU, 2019.

[11] Ginetta Fassio and Christian Simon. "Optimierung audiovisueller Medien für ein hörgeschädigtes Publikum". Diploma Thesis. Hochschule für Film und Fernsehen Potsdam Babelsberg, 2010.

[12] Lionel Fontan et al. "Relationship Between Speech Intelligibility and Speech Comprehension in Babble Noise". In: *Journal of Speech, Language, and Hearing Research* 58.3 (June 2015). ISSN: 1092-4388. DOI: 10.1044/2015_JSLHR-H-13-0335.

[13] Harald Fuchs and Dirk Oetting. "Advanced clean audio solution: Dialogue enhancement". In: *SMPTE Motion Imaging Journal* 123.5 (2014). ISSN: 21602492. DOI: 10.5594/j18429.

[14] Hubert Gabrisch et al. *Alter im Wandel - Ältere Menschen in Deutschland und der EU*. Tech. rep. 7-8. Wiesbaden: Statistisches Bundesamt, 2012.

[15] Yannik Grewe, Christian Simon, and Ulli Scuda. "Producing Next Generation Audio using the MPEG-H TV Audio System". In: *Proceedings of Broadcast Engineering and Information Technology Conference* (2018).

[16] Jürgen Herre et al. "MPEG-H 3D Audio - The New Standard for Coding of Immersive Spatial Audio". In: *IEEE Journal on Selected Topics in Signal Processing* 9.5 (2015). ISSN: 19324553. DOI: 10.1109/JSTSP.2015.2411578.

[17] Gerhard Hesse and Armin Laubert. "Hörminderung im Alter–Ausprägung und Lokalisation". In: *Deutsches Ärzteblatt* 102.42 (2005).

[18] Elisabeth Hildebrandt. "Sprachverständlichkeit im Fernsehen". Diploma Thesis. Universität für Musik und darstellende Kunst Wien, 2014.

[19] iZotope Inc. *iZotope - RX7*. 2019. URL: https://www.izotope.com/en/products/rx.html (visited on 11/20/2019).

[20] IRT GmbH. *Object-based audio - The future of audio production, delivery and consumption*. 2020. URL: https://lab.irt.de/demos/object-based-audio/ (visited on 03/10/2020).

[21] ITU. *Algorithms to measure audio programme loudness and true-peak audio level - ITU-R BS.1770-4*. Tech. rep. Geneva, 2017.

[22] ITU. *Making Television Accessible*. Tech. rep. Geneva: ITU, 2011.

[23] Sofie Jansen et al. "Comparison of three types of French speech-in-noise tests: A multi-center study". In: *International Journal of Audiology* 51.3 (2012). DOI: 10.3109/14992027.2011.633568.

[24] Jean-Marc Jot, Brandon Smith, and Jeffrey Thompson. "Dialog control and enhancement in object-based audio systems". In: *139th Audio Engineering Society International Convention, AES 2015* 139 (2015).

[25] Cheol Yong Kang et al. "Listener Auditory Perception Enhancement using Virtual Sound Source Design for 3D Auditory System". In: *International journal of advanced smart convergence* 5.4 (2016). ISSN: 2288-2847. DOI: 10.7236/ijasc.2016.5.4.15.

[26] K. Kjörling et al. "AC-4 - The next generation audio codec". In: *140th Audio Engineering Society International Convention* (2016).

[27] Tomoyasu Komori et al. "An investigation of audio balance for elderly listeners using loudness as the main parameter". In: *125th Audio Engineering Society Convention* 1 (2008).

[28] Jan Löhler et al. "Schwerhörigkeit im Alter – Erkennung, Behandlung und assoziierte Risiken". In: *Deutsches Ärzteblatt* 116.17 (2019). ISSN: 18660452. DOI: 10.3238/arztebl.2019.0301.

[29] C. D. Mathers. *A study of sound balances for the hard of hearing - Research Department Report*. Tech. rep. BBC, 1991.

[30] Stefan Meltzer et al. "MPEG-H 3D Audio - The Next Generation Audio System". In: *International Broadcasting Convention (IBC) 2014 Conference*. Amsterdam: Institution of Engineering and Technology, 2014. ISBN: 978-1-84919-927-8. DOI: 10.1049/ib.2014.0011.

[31] Nicole Menche, ed. *Biologie Anatomie Physiologie*. 7th editio. Munich: Urban & Fischer bei Elsevier, 2012. ISBN: 9783437268021.

[32] Brian C. J. Moore et al. *Basic Aspects of Hearing*. Ed. by Brian C J Moore et al. 1st editio. Vol. 787. New York: Springer Science+Business Media, 2013. ISBN: 978-1-4614-1590-9. DOI: 10.1007/978-1-4614-1590-9.

[33] Manuel Moussallam. *Releasing Spleeter: Deezer Research source separation engine*. 2019. URL: https://deezer.io/releasing-spleeter-deezer-r-d-source-separation-engine-2b88985e797e (visited on 11/20/2019).

[34] Hannes Münsch. "Aging and sound perception: Desirable characteristics of entertainment audio for the elderly". In: *125th Audio Engineering Society Convention* 125.10 (2008).

[35] Pilar Orero, Carlos Alberto Martin, and Mikel Zorrilla. "HBB4ALL: Deployment of HbbTV services for all". In: *IEEE International Symposium on Broadband Multimedia Systems and Broadcasting, BMSB* (2015). ISSN: 21555052. DOI: 10.1109/BMSB.2015.7177252.

[36] Jouni Paulus et al. "MPEG-D spatial audio object coding for dialogue enhancement (SAOC-DE)". In: *138th Audio Engineering Society Convention* 1 (2015).

[37] Jouni Paulus et al. "Source Separation for Enabling Dialogue Enhancement in Object-based Broadcast with MPEG-H". In: *Journal of the Audio Engineering Society* 67.7/8 (2019). ISSN: 15494950. DOI: 10.17743/jaes.2019.0032.

[38] Samsung. *Dialogue Clarity - Samsung TV*. 2018. URL: https://www.samsung.com/in/support/tv-audio-video/what-is-dialog-clarity-in-samsung-tv/ (visited on 11/21/2019).

[39] Sennheiser electronic GmbH & Co. KG. *The new Sennheiser RS 195 - Rediscover the pleasures of listening*. 2020. URL: https://en-de.sennheiser.com/audiophile-headphones-wireless-digital-over-ear-rs-195 (visited on 03/08/2020).

[40] Ben Shirley and Rob Oldfield. "Clean audio for TV broadcast: An object-based approach for hearing-impaired viewers". In: *Journal of the Audio Engineering Society* 63.4 (2015). ISSN: 15494950. DOI: 10.17743/jaes.2015.0017.

[41] Benjamin Shirley and Paul Kendrick. "ITC Clean Audio Project". In: *116th Audio Engineering Society Convention* (2004).

[42] Ben Shirley et al. "Personalized object-based audio for hearing impaired TV viewers". In: *Journal of the Audio Engineering Society* 65.4 (2017). ISSN: 15494950. DOI: 10.17743/jaes.2017.0005.

[43] Christian Simon, Matteo Torcoli, and Jouni Paulus. "MPEG-H Audio for Improving Accessibility in Broadcasting and Streaming". 2019. URL: http://arxiv.org/abs/1909.11549.

[44] SONOS. *Sprachverbesserung und Nachtmodus für deine Playbar, Playbase, und Beam*. 2019. URL: https://support.sonos.com/s/article/4796?language=en_US (visited on 11/21/2019).

[45] Statistisches Bundesamt. *Ältere Menschen in Deutschland und der EU*. Tech. rep. Wiesbaden: Statistisches Bundesamt, 2016. URL: https://www.destatis.de/DE/Themen/Gesellschaft-Umwelt/Bevoelkerung/Bevoelkerungsstand/Publikationen/Downloads-Bevoelkerungsstand/broschuere-aeltere-menschen-0010020169004.pdf?__blob=publicationFile.

[46] Statistisches Bundesamt Wiesbaden. *Bevölkerung im Wandel - Annahmen und Ergebnisse der 14. koordinierten Bevölkerungsvorausberechnung*. Tech. rep. Wiesbaden: Statistisches Bundesamt, 2019.

[47] Steinberg Media Technologies GmbH. *Premium Audio for Professionals - Nuendo*. 2020. URL: https://new.steinberg.net/nuendo/ (visited on 02/28/2020).

[48] Michael Streppel et al. *Hörstörungen und Tinnitus- Gesundheitsberichterstattung des Bundes*. Tech. rep. Berlin: Robert Koch-Institut, 2006.

[49] The Mathworks Inc. *MATLAB*. 2020. URL: https://www.mathworks.com/products/matlab.html (visited on 02/28/2020).

[50] Matteo Torcoli et al. "Background Ducking to Produce Esthetically Pleasing Audio for TV with Clear Speech". In: *146th Audio Engineering Society Convention* (2019).

[51] Matteo Torcoli et al. "Preferred Levels for Background Ducking to Produce Esthetically Pleasing Audio for TV with Clear Speech". In: *Journal of the Audio Engineering Society* 67.12 (Dec. 2019). ISSN: 15494950. DOI: 10.17743/jaes.2019.0052.

[52] Matteo Torcoli et al. "The Adjustment / Satisfaction Test (A/ST) for the Subjective Evaluation of Dialogue Enhancement". In: *143rd Audio Engineering Society Convention* (2017). URL: http://www.aes.org/e-lib/browse.cfm?elib=19239.

[53] Matteo Torcoli et al. "The Adjustment/Satisfaction Test (A/ST) for the evaluation of personalization in broadcast services and its application to dialogue enhancement". In: *IEEE Transactions on Broadcasting* 64.2 (2018). ISSN: 00189316. DOI: 10.1109/TBC.2018.2832458.

[54] Emmanuel Vincent. *Blind Audio Source Separation A review of state-of-the-art techniques*. London, 2005.

[55] Bianka Wachtlin. "Überblick über Erkrankungen im Alter, die zum Hörverlust führen und die Versorgung mit Hilfsmitteln". In: *Fachtagung "Nicht nur sehbehindert?" Der VBS-AG Rehabilitation und gresellschaftliche Teilhabe sehbehinderter und blinder Seniorinnen und Senioren*. February. Marburg a.d. Lahn, 2017.

[56] Lauren Ward. "Improving broadcast accessibility for hard of hearing individuals - using object-based audio personalisation and narrative importance". PhD. University of Salford, 2019.

[57] Lauren Ward, Ben Shirley, and Jon Francombe. "Accessible object-based audio using hierarchical narrative importance metadata". In: *145th Audio Engineering Society Convention* (2018).

[58] Lauren Ward et al. "Casualty Accessible and Enhanced (A&E) Audio: Trialling object-based accessible TV audio". In: *147th Audio Engineering Society Convention* (2019).

[59] World Health Organization. *Addressing the rising prevalence of hearing loss*. Tech. rep. Geneva: World Health Organization, Aug. 2018. URL: https://apps.who.int/iris/handle/10665/260336.

[60] World Health Organization. *Prevention of blindness and deafness - Grades of hearing impairment*. 2020. URL: https://www.who.int/pbd/deafness/hearing_impairment_grades/en/ (visited on 01/23/2020).

[61] Thomas Zahnert. "The Differential Diagnosis of Hearing Loss". In: *Deutsches Arzteblatt* 108.25 (2011). ISSN: 00121207. DOI: 10.3238/arztebl.2011.0433.

# A. German version of the A/ST

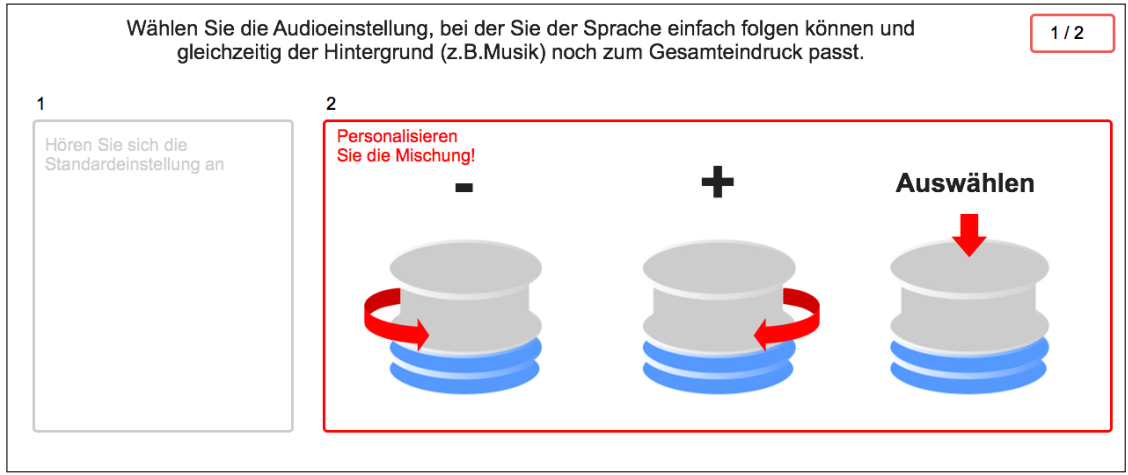

GUI of the LD adjustment phase in the A/ST as it appeared in the test in German.

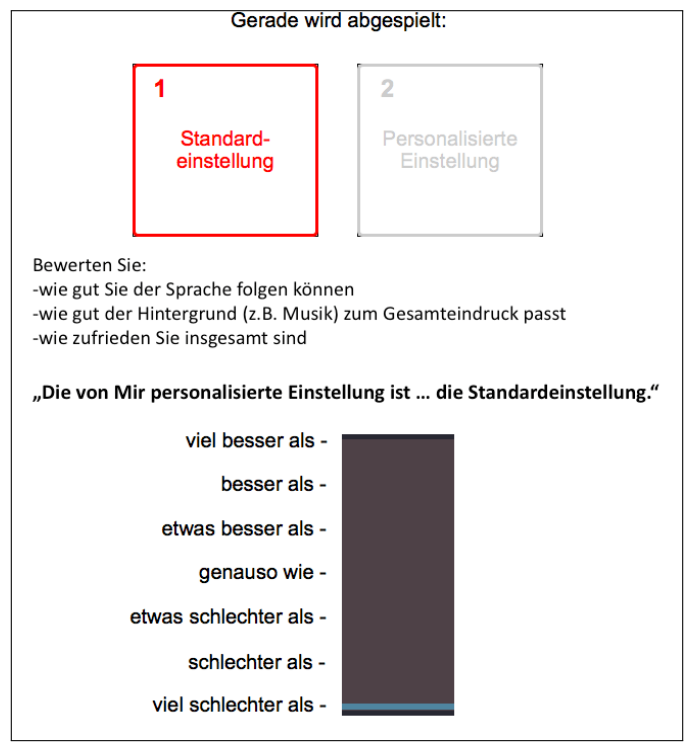

GUI of the adjustment satisfaction phase in the A/ST as it appeared in the test in German.

B QUESTIONNAIRE

# B. Questionnaire

© Fraunhofer IIS – February 2020                                             Davide Straninger v3 GER

Testteilnehmer ID:\_\_\_\_\_\_

**Qualitative Fragen zum Hörtest** (A/ST)

Beantworten Sie bitte die folgenden Fragen **kurz und bündig** in eigenen Worten. Stichpunkte sind erlaubt.

**0. Wie schätzen Sie Ihr eigenes Hörvermögen ein?**
(Zutreffendes bitte ankreuzen!)

- ○ Ausgezeichnet
- ○ Gut
- ○ Durchschnittlich
- ○ Mäßig
- ○ Schlecht

**1. Wie gut war die Benutzeroberfläche zu verstehen und zu benutzen?**

**2. Hat Ihnen die Möglichkeit gefallen, die Tonmischungen zu personalisieren?**

**3. Was war Ihrer Meinung nach der Zweck dieses Tests?**

First page of the questionnaire in German.

**4. Aufgrund welcher Faktoren haben Sie Ihre Bewertungen getroffen?**

**5. Wie oft haben Sie Probleme damit, die Sprache im TV zu verstehen?**
(Zutreffendes bitte ankreuzen!)

- Jeden Tag
- Mindestens 1x in der Woche
- Mindestens 1x im Monat
- Niemals

**6. Haben Sie allgemeine Anmerkungen zum Test?**

**Vielen Dank für Ihre Teilnahme!**

Second page of the questionnaire in German.

# C. Further information on the feasibility study

To be able to vary the LD between FG and BG there have to be single steps in which the participants change it. During testing it became evident that the steps should not be too big, because otherwise the differences between the individual gradations would be audible. So the steps in which the LD can be changed had to be small enough so that no disturbing loudness jumps could be heard when turning the rotary knob, but at the same time large enough so that no unnecessary amount of data were generated. Further information on the items used can be found in chapter 4.4.

In a first draft of the test, the BG of the test items was a dynamically ducked[13] BG that could then be attenuated or amplified in loudness by turning the knob. This is a common way of mixing the BG loudness of a TV program in order to create a more clear speech signal. After some tests it became clear that small adjustments of the LD between a FG signal and a dynamically ducked BG is hardly audible even for professionals and hence not suitable for listeners who are hard of hearing. Therefore it was decided to stick to a statically attenuated[14] BG that could be changed in loudness.

During the search for suitable test items, the idea of using original television material came up. The "Tatort", a crime series and one of the most popular TV shows in Germany, was a good choice for this purpose. Due to the tension to be created in the film, many scenes with low LDs appear. Often there are scenes with speech and simultaneous noise or music. Also the "Tatort" is one of the programs in German television which is often complained about when it comes to speech intelligibility. For this work the episode "Tiefer Sturz" was kindly provided by the German broadcaster WDR.

It was also thought about the simultaneous presentation of the corresponding video to the audio material. While the presentation of a visual stimulus would have certainly been closer to reality, it could have distracted the subjects from the already difficult task of detecting small changes in the LD. Especially in the very subconscious area of auditory perception and with older test participants presumably affected by presbycusis, care must be taken to minimize distracting influences. Therefore it was decided to conduct the listening test without video presentation.

Finally, the question arose as to which LD should be taken as the default value with which the personalized version can be compared. A very low SNR e.g. would presumably motivate the test participant to start adjusting the LD. It was decided to use as standard value the LD which was found in the original television material at the time of broadcast, in order to be able to check at the same time whether the test subjects would consider this LD to be acceptable or would prefer to set a different SNR.

---

[13]The term dynamical ducking refers to a time-varying BG attenuation that occurs when a trigger signal, e.g. speech, is present at the same time.

[14]Statical attenuation refers to a BG being time-constantly attenuated in loudness.

# D. Item LD steps

| Item name | Default LD for OO and DS | LD steps |
|---|---|---|
| WDR1 | 11 LU | +12:1:-15;-16:2:-40 |
| WDR2 | 8.2 LU | +12:1:-15;-16:2:-40 |
| WDR3 | 12.1 LU | +12:1:-15;-16:2:-40 |
| WDR4 | 13 LU | +12:1:-15;-16:2:-40 |
| WDR5 | 11.7 LU | +12:1:-15;-16:2:-40 |
| AR1 | 4 LU | +9.6:0.2:0;-0.8:0.8:-20 |
| AR2 | 1 LU | +9.6:0.2:0;-0.8:0.8:-20 |
| AR3 | 6 LU | +9.6:0.2:0;-0.8:0.8:-20 |

Default LDs and LD steps of the items according to the naming in the plots.
LD step naming follows the scheme: From $\Delta$ xx LU : in x LU steps : to $\Delta$ xx LU; from $\Delta$ xx LU : in x LU steps : to $\Delta$ xx LU.

# E. Listening test instructions

© Fraunhofer IIS - February 2020                                                    Davide Straninger v4 GER

## Hörtest zur Evaluierung der Hörerzufriedenheit
- Anleitung für Testteilnehmende -

**Überblick**

In diesem Hörtest soll Ihre Präferenz zum Lautstärkeunterschied zwischen der Sprache und dem Hintergrund (Musik, Audioeffekte, etc.) in Fernsehprogrammen untersucht werden. Der Test gliedert sich in vier Teile:

**1. Gesamtlautstärke einstellen**

Bitte stellen Sie als erstes die Gesamtlautstärke für den Test so ein, wie Sie zu Hause fernsehen würden. Diese Lautstärke darf danach nicht mehr verändert werden.

**2. Tonmischung personalisieren**

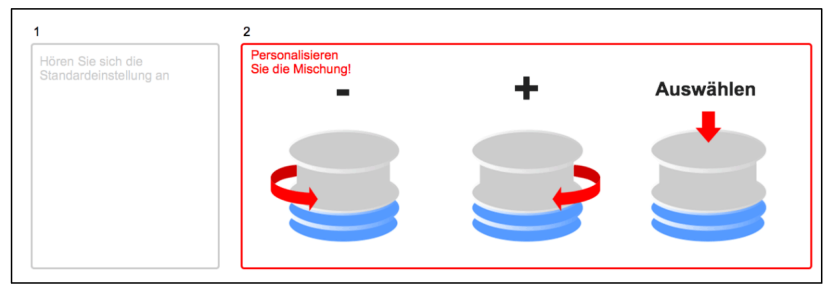

Sie werden im Folgenden verschiedene Tonausschnitte aus Fernsehprogrammen hören. Drehen Sie bitte am silbernen Drehknopf, um den Lautstärkeunterschied zwischen Sprache und Hintergrund (Musik, Audioeffekte, etc.) so zu verändern, dass Sie dem Dialog gut folgen können und der Hintergrund gleichzeitig noch zum Gesamteindruck passt.
Durch Ihre Personalisierung kann sich die Audioqualität verschlechtern. Finden Sie in diesem Fall bitte einen guten Mittelwert zwischen Sprachverständlichkeit und Audioqualität, den Sie auch im Fernsehen akzeptieren würden.
Um den Unterschied zwischen der Standardeinstellung und Ihrer personalisierten Einstellung zu hören, können Sie die **Tasten 1 und 2** auf der Tastatur benutzen. Wenn Sie eine Einstellung gefunden haben die Ihnen gefällt, drücken Sie den silbernen Knopf.

Page 1 of the instructional sheet for the listening test in German.

© Fraunhofer IIS - February 2020                                          Davide Straninger v4 GER

**3.** Personalisierte Tonmischung bewerten

Gerade wird abgespielt:

**1** Standard-einstellung

2 Personalisierte Einstellung

Bewerten Sie:
- wie gut Sie der Sprache folgen können
- wie gut der Hintergrund (z.B. Musik) zum Gesamteindruck passt
- wie zufrieden Sie insgesamt sind

„Die von Mir personalisierte Einstellung ist … die Standardeinstellung."

viel besser als -
besser als -
etwas besser als -
genauso wie -
etwas schlechter als -
schlechter als -
viel schlechter als -

Es erscheint ein neues Fenster auf dem Bildschirm. Benutzen Sie die **Tasten 1 und 2** auf der Tastatur um zwischen der „Standardeinstellung" und Ihrer „Personalisierten Einstellung" zu wechseln. Hören Sie auf die Unterschiede der beiden Versionen. Bewerten Sie die von Ihnen personalisierte Version des Tonausschnittes entsprechend der Beschreibung auf dem Bildschirm. Drehen Sie dazu bitte den silbernen Knopf. Haben Sie Ihre Bewertung getroffen, drücken Sie den Knopf.

**4. Nächster Tonausschnitt**
Nun folgen die **Schritte 2 und 3** erneut. Diesmal nur mit einem anderen Tonausschnitt. Führen Sie den Test bitte so lange durch, bis ein grünes Fenster mit dem Schriftzug „Vielen Dank!" zu sehen ist. Damit ist der Hörtest beendet. Anschließend werden Ihnen Fragen zum Test gestellt. Wenn Sie zwischendurch eine Hörpause benötigen, können Sie den Ton mit der Leertaste pausieren oder weiterlaufen lassen. Sollten Sie noch Fragen haben, zögern Sie bitte nicht, diese zu stellen, denn es ist wichtig, dass Sie den Testablauf verstanden haben.

Vielen Dank für Ihre Teilnahme!

Page 2 of the instructional sheet for the listening test in German.

# F. Participant while doing the listening test

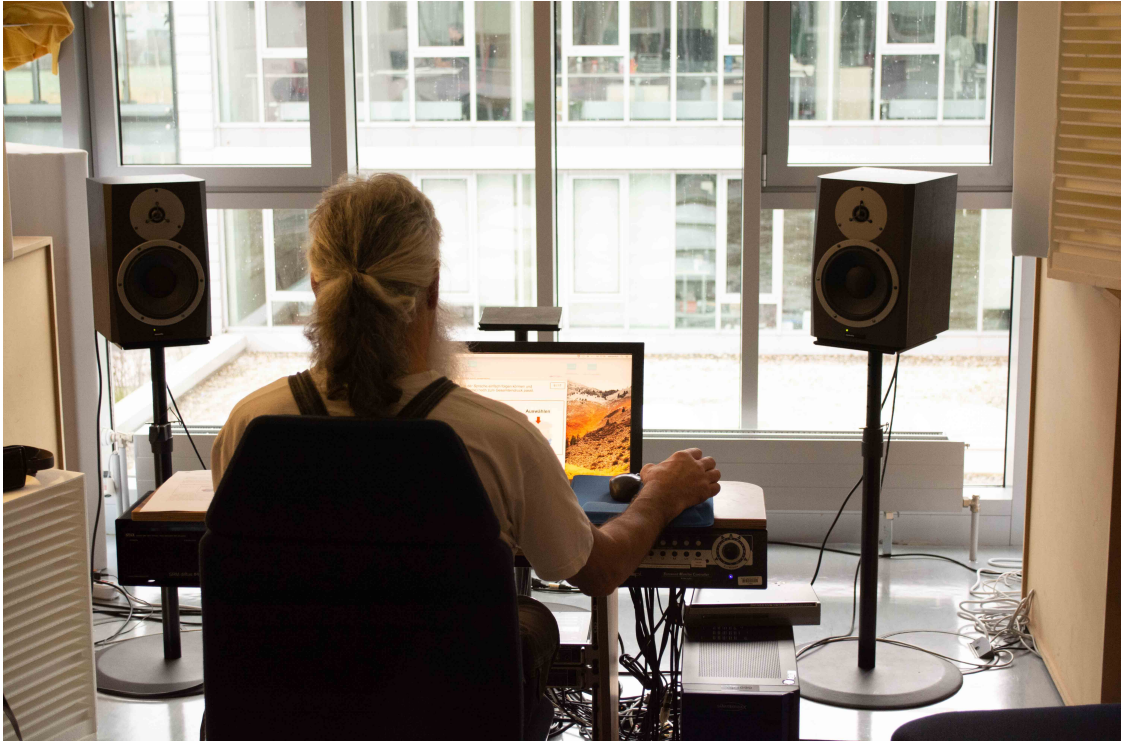

Picture of a test participant while working on the A/ST in the listening test room.

# G. Raw data on personalization in the A/ST

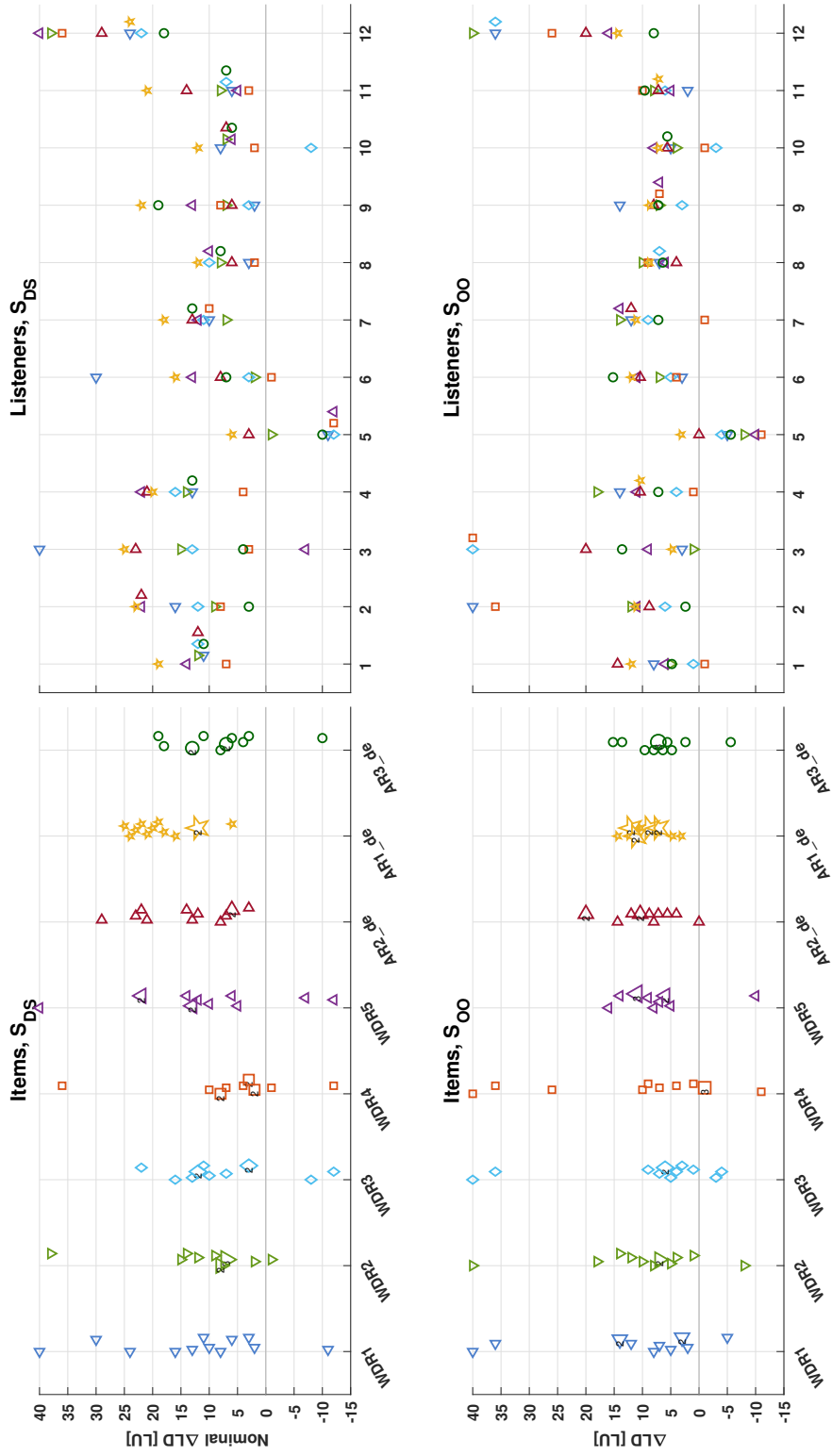

Raw data on the preferred LD for items and for listeners in relative LD values.

# H. Raw data on satisfaction assessment in the A/ST

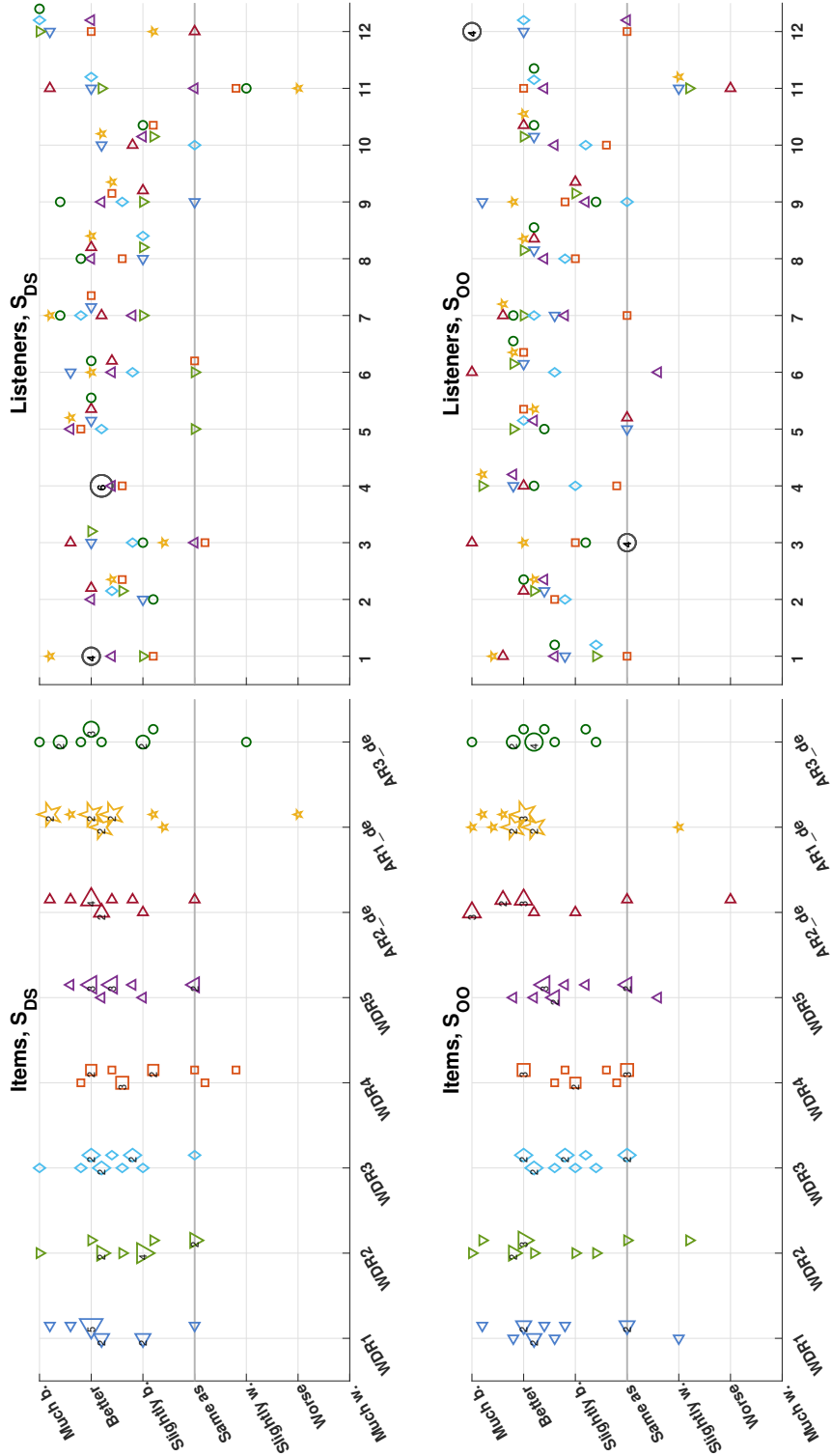

Raw data on the satisfaction assessment for items and for listeners in relative LD values.